\documentclass{aa}
\pdfoutput=1
\usepackage{graphicx}
\usepackage{txfonts}
\usepackage{xcolor, textcomp}
\usepackage{enumitem}
\newcommand{\ts}{\textsuperscript}
\newcommand\T{\rule{0pt}{2.8ex}}       % Top strut
\newcommand\B{\rule[-1.4ex]{0pt}{0pt}}       % Bottom strut
\renewcommand{\bold}[1]{\normalfont{#1}}
\begin{document} 

   \title{Parameter optimization for surface flux transport models}

   \author{T. Whitbread\inst{1}
          \and
          A.~R. Yeates\inst{1}
          \and
          A. Mu\~noz-Jaramillo\inst{2}
          \and
          G.~J.~D. Petrie\inst{3}
          }

   \institute{Department of Mathematical Sciences, Durham University, Durham, DH1 3LE, UK\\
              \email{tim.j.whitbread@durham.ac.uk}
         \and
             Southwest Research Institute, 1050 Walnut St. \#300, Boulder, CO 80302, USA\\
             \email{amunozj@boulder.swri.edu}
         \and
             National Solar Observatory, Boulder, CO 80303, USA\\
             \email{gpetrie@nso.edu}}         

   \date{Received ; Accepted }

\abstract{Accurate prediction of solar activity calls for precise calibration of solar cycle models. Consequently we aim to find optimal parameters for models which describe the physical processes on the solar surface, which in turn act as proxies for what occurs in the interior and provide source terms for coronal models. We use a genetic algorithm to optimize surface flux transport models using National Solar Observatory (NSO) magnetogram data for Solar Cycle 23. This is applied to both a 1D model that inserts new magnetic flux in the form of idealized bipolar magnetic regions, and also to a 2D model that assimilates specific shapes of real active regions. The genetic algorithm searches for parameter sets (meridional flow speed and profile, supergranular diffusivity, initial magnetic field, and radial decay time) that produce the best fit between observed and simulated butterfly diagrams, weighted by a latitude-dependent error structure which reflects uncertainty in observations. Due to the easily adaptable nature of the 2D model, the optimization process is repeated for Cycles 21, 22, and 24 in order to analyse cycle-to-cycle variation of the optimal solution. We find that the ranges and optimal solutions for the various regimes are in reasonable agreement with results from the literature, both theoretical and observational. The optimal meridional flow profiles for each regime are almost entirely within observational bounds determined by magnetic feature tracking, with the 2D model being able to accommodate the mean observed profile more successfully. Differences between models appear to be important in deciding values for the diffusive and decay terms. In like fashion, differences in the behaviours of different solar cycles lead to contrasts in parameters defining the meridional flow and initial field strength.}


\maketitle

\section{Introduction}

Surface flux transport (SFT) is a crucial component of the 11-year sunspot cycle. SFT models have been developed and used for decades \citep[e.g.,][]{devore84,wangetal89,vanballe98,schrijveretal01,baumann04,jiang10} with some success, though results can be sensitive to the choice of parameters.
\\\\
Surface flux transport models track the evolution of magnetic regions, which appear on the solar surface due to the rise of buoyant flux tubes \citep{fan09}. Generally they emerge as bipolar magnetic regions (BMRs) with a leading polarity and a trailing polarity with respect to the east-west direction. The leading polarities during a single cycle are usually the same across a hemisphere, with the reverse pattern occurring on the opposite hemisphere, following Hale's polarity law \citep{hale}. Furthermore, the polarities reverse at the end of each cycle, giving rise to a magnetic cycle with a period of approximately 22 years. At the start of a cycle, magnetic regions emerge at latitudes of approximately $\pm 30$\textdegree{}. As the cycle progresses, emergence regions migrate equatorwards before reaching $\pm 5$\textdegree{} latitude at the end of the cycle. This emergence pattern was noted by \citet{sporer}, and can be observed in the well-known `butterfly diagram'.
\\\\
Bipolar magnetic regions tend to emerge tilted at an angle with respect to the east-west line (using the line connecting the centres of the opposing polarities), with the leading polarity emerging closer to the equator. This effect is due to the helical convective motions in the convection zone and is more pronounced at higher latitudes, according to Joy's law \citep{joyslaw}. In addition, the granular and supergranular convection cells provide a means of diffusion whereby flux is transported to the edge of the convection cells, spreading out across the solar surface and resulting in the leading flux cancelling across the equator with the corresponding opposite flux \citep{leighton64}. The remaining trailing flux is transported to the polar regions via a combination of turbulent diffusion and a meridional flow \citep{howard79}. Meridional circulation is a relatively slow transport mechanism, with speeds of $\sim$\,10--20\,m\,s$^{-1}$ observed via helioseismology by e.g., \citet{braun98}, \citet{zhao04}, \citet{jackie}. However, helioseismic recordings are near the limit of credibility \citep{komm13} and so the flow profile is not well constrained. Flow speeds within this range have also been found by e.g., \citet{komm93} via the tracking of small magnetic features, and by \citet{topka} via the comparison of polar zone filament distribution and polar magnetic field evolution. When the remaining flux eventually reaches the pole it cancels with the polar field from the previous cycle, culminating in polar field reversal. The reversal typically occurs at cycle maximum. For a historical review of surface flux transport, see \citet{sheeleyreview}.
\\\\
There is evidence to suggest that the strength of the polar field at the end of the cycle is a good indicator of the strength of the following cycle \citep[e.g.,][]{schatten78,andres13}. Moreover, the strength of a cycle is anti-correlated with the duration of the cycle, according to the Waldmeier Effect \citep{waldmeier}. An explanation is offered by \citet{cameron16} whereby activity belts near the base of the convection zone cancel across the equator with the opposing polarity. Stronger cycles have wider belts and so cancellation occurs earlier, resulting in a shorter cycle since all declining phases are approximately the same.
\\\\
There is an ever-present need to predict the amplitude and length of future solar cycles since solar weather can have adverse effects on satellites, astronauts and technological systems on Earth. Given that disturbances such as coronal mass ejections and solar flares are usually attributed to regions of strong magnetic field, tracking and modelling magnetic regions on the solar surface, i.e., surface flux transport, has been highlighted as a key method for predicting and understanding solar cycle variability \citep[e.g.,][]{hathuptonpredict,hathupton16}, without using the more complicated calculations involved in dynamo simulations \citep[for a review of dynamo theory, see][]{charreview}.
\\\\
As mentioned above, the output of current SFT models is highly dependent on the parameters used. Parametrizations of the transport processes have been made, particularly for diffusion and meridional flow, but the exact forms are still debated, and are not necessarily in line with the limited observations available. Parameter studies of varying scope have been undertaken \citep[e.g.,][]{schrijver02,durrant03,baumann04,yeates14}, but without complete parameter coverage. In this paper we aim to systematically find optimal parameters to be used in SFT models which accurately reproduce such features of the solar cycle as poleward flux transport `surges', polar field reversal time, and polar field strength. The results can also be used to constrain the surface components of dynamo simulations to produce the most accurate cycle predictions to date. A similar study was performed by \citet{lemerle}, who used the same genetic algorithm used in this paper to find optimal parameters for a 2D SFT model for Cycle 21 only, with the view of coupling it to a 2D flux transport dynamo model \citep{lemerle2}. In contrast we analyse two distinct models with different dimensionality, namely 1D and 2D, and different data-assimilation techniques, initially for Cycle 23. While \citet{lemerle} used the comprehensive BMR database compiled by \citet{wangsh89}, we use the same BMR database as \citet{yeates07} for the 1D model, and extract individual active regions from synoptic magnetograms for the 2D model. We also apply the 2D model to other cycles to search for cyclical variation.
\\\\
In Sect. \ref{sect2} we present the 1D model and the genetic algorithm used to perform the optimization, including a prescribed error structure dependent on latitude and magnetic field strength to factor in observational uncertainty. We also discuss the results of the optimization for Cycle 23. In Sect. \ref{sect3} we present the 2D model which directly assimilates active regions into the simulation, and run the optimization process on this model for Cycle 23. In Sect. \ref{sect4} we compare our optimal meridional flow profiles from both models with observations. Finally, we perform optimizations on the 2D model for Cycles 21, 22, and 24 in Sect. \ref{sect5}, before concluding in Sect. \ref{conclusions}.

\section{One-dimensional model} \label{sect2}

Since we aim to optimize the parameters of global, axisymmetric flows using the longitude-averaged butterfly diagram, we start by considering a 1D SFT model. Using idealized BMRs as input, this model has the advantage of allowing many realizations to be run rapidly. In Sect. \ref{sect3}, we will compare the parameters selected by the 1D model to a 2D model with more realistic assimilation of active region data.

\subsection{Numerical method} \label{sect21}

\begin{figure}
  \resizebox{\hsize}{!}{\includegraphics{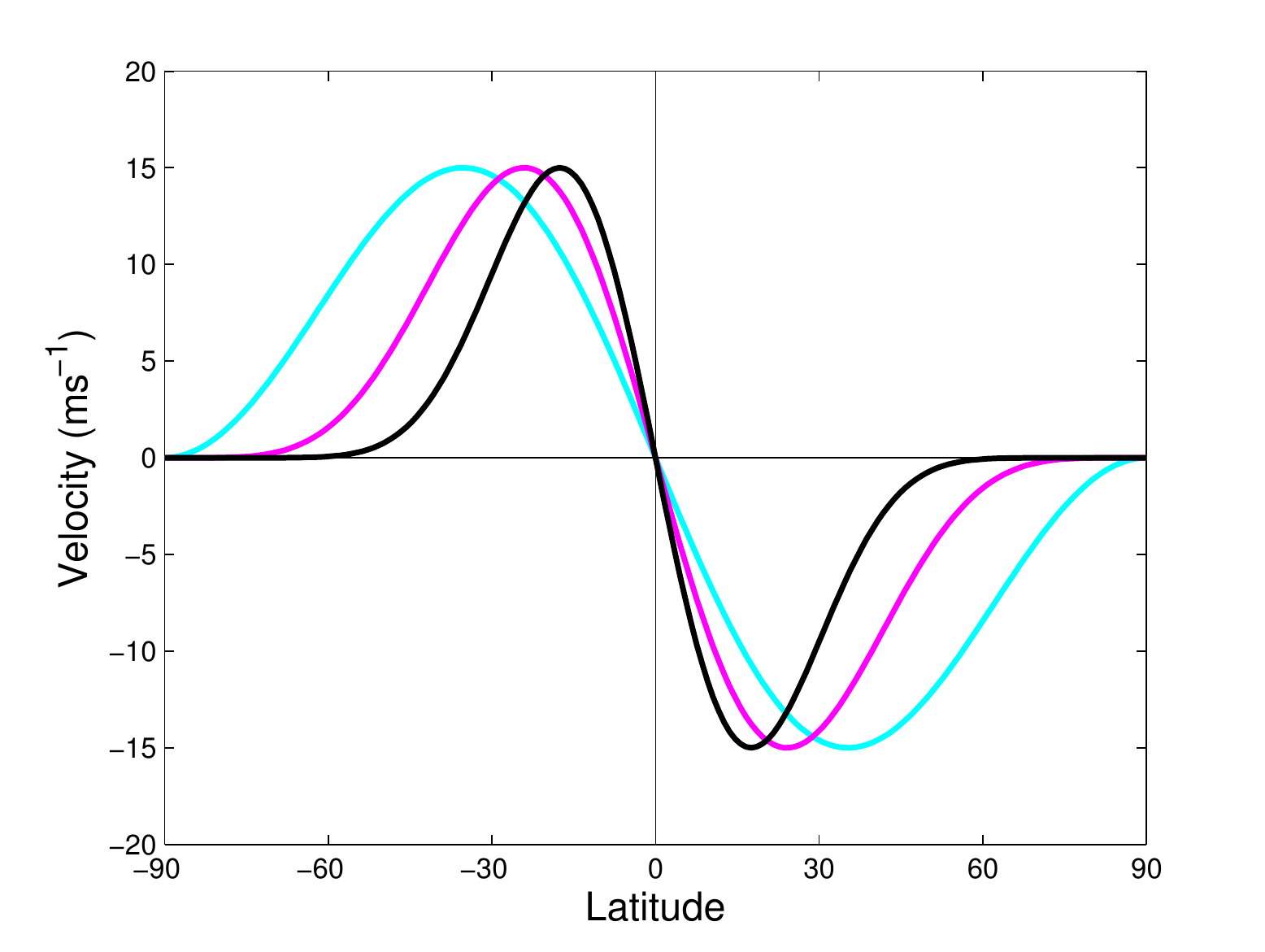}}
  \caption{Three examples of the meridional flow profile in Equation \ref{meridfloweqn} for $v_0 = 15$\,m\,s$^{-1}$, $p=2$ (cyan), $p=5$ (magenta) and $p=10$ (black).}
  \label{meridflow}
\end{figure}

In the SFT model, the radial component of the magnetic field after averaging in longitude, $B\left(\theta ,t\right)$, evolves according to the advection-diffusion equation \citep{leighton64}:
\begin{eqnarray} \label{sfteqn1d}
  \frac{\partial{B}}{\partial{t}} &=& \frac{\eta}{R_{\odot}^2\sin\theta}\frac{\partial}{\partial{\theta}}\left(\sin\theta\,\frac{\partial{B}}{\partial{\theta}}\right) - \frac{1}{R_{\odot}\sin\theta}\frac{\partial}{\partial{\theta}}\Big(v\left(\theta\right)\sin\theta\,B\Big)\nonumber\\
  &&-\> \frac{1}{\tau}B + S(\theta,t) ,
\end{eqnarray}
where $R_{\odot}$ is solar radius, $\eta$ is diffusivity, representing the diffusive effect of granular convective motions, $\tau$ is an exponential decay term added by \citet{schrijver02} to improve regular polar field reversal, and $S$ is a source term for newly emerging magnetic regions. The profile $v\left(\theta\right)$ describes poleward meridional flow. The magnetic field is decomposed into spherical harmonic form:
\begin{equation}
  B\left(\theta,t\right) = \sum_{l=1}^\infty a_{l0}\left(t\right)Y_l^0\left(\theta\right) ,
\end{equation}
and the resulting coupled ordinary differential equations for the coefficients $a_{l0}\left(t\right)$ are solved using an adaptive Runge-Kutta (4,5) time-stepping method \citep{rk45}. The equations are solved on a grid of 180 cells equally spaced in latitude.
\\\\
In the 1D model, new magnetic regions are assumed to take the form of idealized BMRs. \bold{Each BMR has a specified day of emergence; longitude and latitude; size; magnetic flux, including polarity; and tilt angle taken from an existing observational dataset where these properties were determined individually for each BMR from NSO synoptic magnetograms \citep{yeates07}}. Using these data, the 1644 recorded BMRs from Cycle 23 (1\ts{st} June 1996--3\ts{rd} August 2008) are averaged in longitude and inserted into the model on the corresponding days of emergence. The BMR data \bold{are freely available} at the Solar Dynamo Dataverse\footnotemark \citep{bmrdata}.
\footnotetext{https://dataverse.harvard.edu/dataverse/solardynamo}
\\\\
Because the meridional flow is a relatively slow transport mechanism, there is still a fair amount of uncertainty regarding its properties.  A flexible profile is chosen to factor observational uncertainty into the optimization:
\begin{equation}\label{meridfloweqn}
  v\left(\theta\right) = -v_0\sin^p\theta\cos\theta ,
\end{equation}
where increasing $p$ produces a steeper gradient at low latitudes and a peak closer to the equator (Fig. \ref{meridflow}). The amplitude $v_0$ is chosen to be the maximum of $|v|$, so increasing $v_0$ increases the height of the peak. Both $v_0$ and $p$ are left as free parameters to be optimized. \citet{lemerle} used a similar, but more sophisticated profile, which is discussed in Sect. \ref{sectcycle21}. This provides substantially more flexibility, but introduces extra parameters into the optimization runs which could hinder convergence to a global maximum. A combination of exponential and sinusoidal functions adapted to helioseismic observations \citep{gizon04} was utilized by \citet{schussler06} for the meridional flow profile. Although this may better represent the actual meridional circulation, it cannot be accurately reproduced by the functional form in Equation \ref{meridfloweqn}. In any case, the true functional form of the observed meridional circulation is uncertain, particularly at high latitudes.\\

\begin{figure}
  \resizebox{\hsize}{!}{\includegraphics{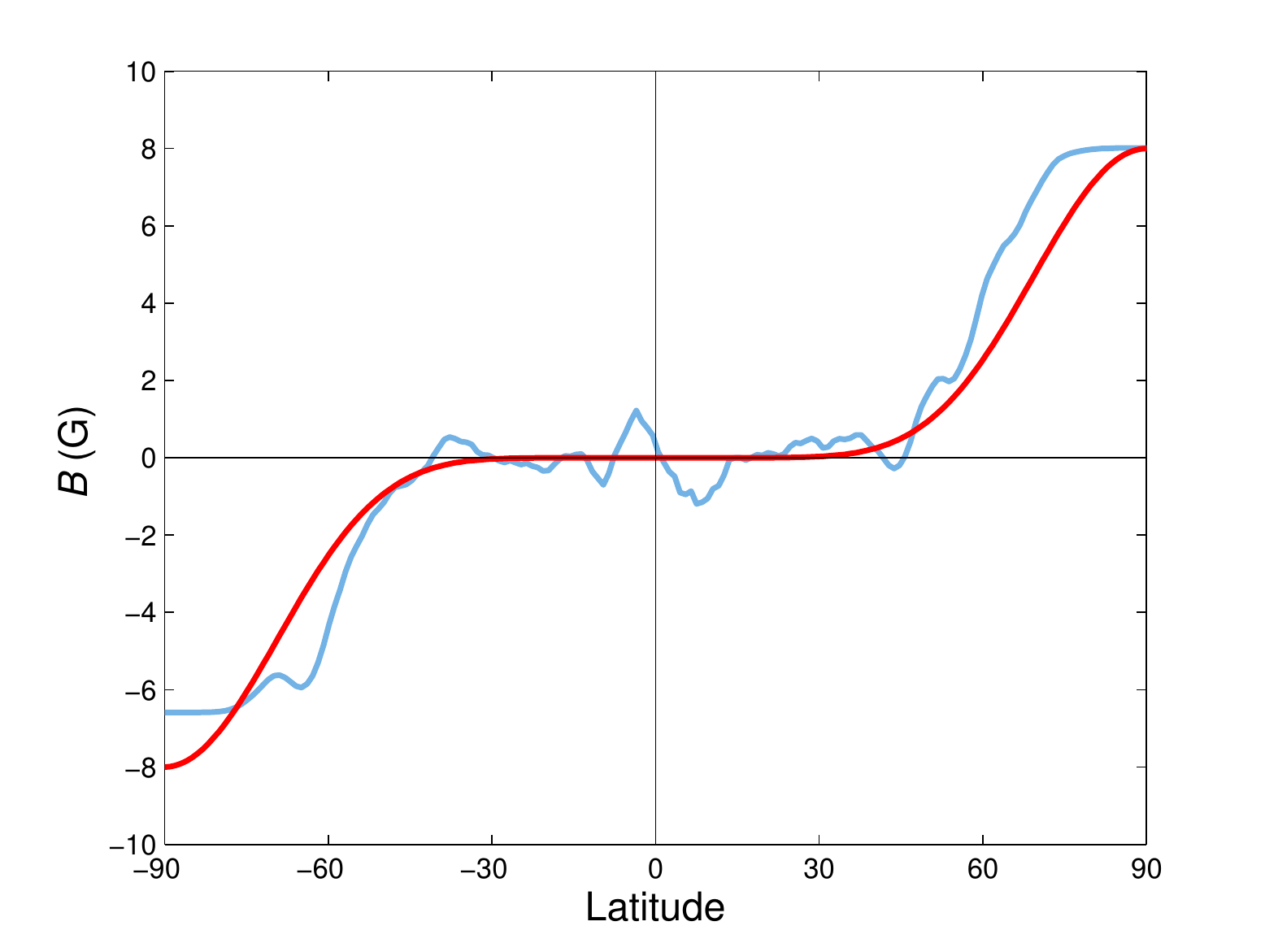}}
  \caption{Comparison between initial magnetogram (blue) and the profile given in Equation \ref{initeqn} (red) with $B_0 = 8$\,G.}
  \label{init}
\end{figure}

\noindent In order to define the initial conditions, we use the profile of \citet{svalgaard78}:
\begin{equation}\label{initeqn}
  B\left(\theta ,0\right) = B_0\,\lvert\cos\theta\rvert ^7 \cos\theta ,
\end{equation}
where $B_0$ is left as a parameter to be optimized. Figure \ref{init} (red) shows a crude fit of this expression to the observed profile from 1910\,CR (blue). Whilst the observed profile is asymmetric across the equator and reveals some activity present at the equator, the prescribed profile represents a typical cycle minimum and ensures that the choice of initial profile is not hindering the optimization process, but rather aiding it with some flexibility. With this form of initial condition we can also be consistent across all regimes. Optimization runs performed using the observed initial condition show that the choice of initial condition in fact has a negligible effect on the optimal butterfly diagram.

\subsection{Ground-truth data} \label{sect22}

As ground-truth data for optimization of the model, we use radial-component magnetogram data from US National Solar Observatory, Kitt Peak, in the form of full-disk images. Prior to 2007\,CR, these came from the Kitt Peak Vacuum Telescope, while from 2007\,CR onwards we use Synoptic Optical Long-term Investigations of the Sun (SOLIS) data\footnotemark\footnotetext{http://solis.nso.edu/0/vsm/vsm\_maps.php}. To minimize noise in the polar regions of the map, we correct the butterfly diagram by calculating a cubic spline interpolation at each latitude of annual average measurements of high-latitude fields (poleward of $\pm 75$\textdegree) which were observed with a preferable solar rotation axis tilt angle. A combination of real and interpolated data is used for the regions between $\pm 60$\textdegree and $\pm 75$\textdegree\ \citep[see][]{petrie12}. The resulting butterfly diagram is interpolated onto a uniform time grid at daily intervals. This is averaged over periods of 27 days, smoothed using a Gaussian filter to bring the unsigned flux down to a comparable level to the simulation, and finally sampled at the resolution of once per Carrington rotation. It should be noted that recently the SOLIS magnetograms were uniformly reprocessed and recalibrated. The butterfly diagram used here was made after this reprocessing.

\subsection{Optimization algorithm}\label{pikaia}

To search for optimal parameter sets where the model matches the observed butterfly diagram, we use the genetic algorithm \texttt{PIKAIA 1.2}. It was written by \citet{pikaia} at the High Altitude Observatory (HAO) of the National Center for Atmospheric Research (NCAR) and is publicly accessible\footnotemark.
\footnotetext{http://www.hao.ucar.edu/modeling/pikaia/pikaia.php}
\\\\
\texttt{PIKAIA} is an evolutionary algorithm originally written in FORTRAN-77. It is particularly effective for multimodal optimization problems, taking advantage of crossover and mutation operators which can induce jumps in parameter space, allowing for greater exploration and reducing the chances of getting trapped at a local maximum which is not the global maximum.
\\\\
The algorithm generates multiple random parameter sets, each entry within a user-defined range, and performs a model simulation for every parameter set, or `population member'. The population are ranked by `fitness' according to a user-defined `fitness function' which, in the case of model optimization, would usually be a comparison between a real reference case map and a model-generated map.
\\\\
The highly ranked population members have a greater chance of being selected for the crossover or `breeding' process whereby sections of corresponding parameter strings from two members are interchanged to produce `offspring', with the aim being that an individual will be produced with desirable features of both `parents' and become the fittest in the population. To increase variability and hence the likelihood for population improvement, random mutation is included, though this is a much slower process than crossover.
\\\\
The crossover-mutation process is run over a user-defined number of generations. Whilst \texttt{PIKAIA} is inherently stochastic and so finding an `acceptable' fit is never guaranteed, a large enough choice for the evolution period should ensure that the combined effect of the crossover and mutation operators has enough time to discover sufficiently fit population members.
\\\\
For a more detailed introduction to \texttt{PIKAIA} and genetic algorithms in general, see \citet{tutorial} and \citet{release}.
\\\\
In order to reduce the duration of computationally intensive optimizations, \citet{mpikaia} created \texttt{MPI\textit{KAIA}}, a freely accessible implementation of \texttt{PIKAIA} 1.2 in MPI\footnotemark 
\footnotetext{http://www.hao.ucar.edu/Public/about/Staff/travis/mpikaia/}. Rather than using a single processor for all model evaluations, a network of computers is used, and each of the model evaluations from a single generation is sent to a separate processor and computed simultaneously, achieving near-perfect parallelization. The time taken to complete the optimization therefore is entirely dependent on the number of generations. For example, for the 1D optimization problem in this paper, each model simulation takes approximately 90 seconds, so, with 46 processors, the runtime is brought down from 24 days to about 12.5 hours for 46 population members evolved over 500 generations. These choices of population size and evolution period should be large enough to obtain good fits to the data, and will be used for the remainder of the paper unless specified.

\subsection{Accounting for uncertainties in observations}

The $\chi^2$ statistic is used as a measure of fit between the real and simulated butterfly diagrams:
\begin{equation}\label{fitness}
  \chi ^2 = \frac{1}{n-k} \sum_{i,j} \left(\frac{B_{obs}\big(\theta _i,t_j\big) - B\big(\theta _i,t_j;\mathbf{x}\big)}{\sigma\left(\theta_i,t_j\right)}\right)^2 ,
\end{equation}
where $n$ is the number of gridpoints and $\mathbf{x}$ is the vector of $k$ free parameters. Since improving best fit is a minimization process and \texttt{PIKAIA} is set up to maximize functions, the reciprocal of the measure, $\chi^{-2}$, is used as the necessary fitness function. The variance $\sigma ^2$ describes the error in both the measurements and the models, and we assume the form:
\begin{equation}
  \sigma ^2\left(\theta_i,t_j\right) = \sigma ^2_{obs}\left(\theta_i,t_j\right) + \sigma ^2_{model} .
\end{equation}
The variance plays two roles in the optimization. Firstly, it gives a meaningful value to the $\chi^{-2}$ statistic. This allows us to compare the performance of different parameter combinations and time periods. Secondly, it effectively weights distinct locations $\left(\theta_i,t_j\right)$ differently in the optimization, since the observed errors are assumed to have the form:
\begin{equation}
  \sigma_{obs}\left(\theta_i,t_j\right) = \frac{0.1\left|B_{obs}\big(\theta_i,t_j\big)\right| + \epsilon}{\sin\theta_i} ,
\end{equation}
where $\epsilon$ is some small increment to ensure that the error is non-zero even in regions of low $B_{obs}$. This error structure reflects the uncertainties and inconsistencies in photospheric magnetic field observations \citep[e.g.,][]{riley14}. The factor of $\sin\theta$ allows for the fact that the errors are in the original line-of-sight measurements, whereas $B_{obs}$ is the inferred radial field \citep{svalgaard78}. Overall, the effect of this error structure reduces the weight of observations both near the pole and in strong active regions. The resulting optimization will favour accuracy in the mid-latitude `transport regions'. A further improvement of the $\chi^{-2}$ statistic would be to include correlation between data points, but for simplicity here we assume independence.
\\\\
Since the model error structure is unknown, we compute $\chi^{-2}$ with $\sigma_{model}=0$. This is sufficient for the purpose of comparing different model runs against the same set of observational data, with a higher value of $\chi^{-2}$ indicating models that give a better match. The simulations are not sufficiently detailed to achieve a significant match at, say, the 99\% level, which is evident from visual inspection of the butterfly diagrams. To achieve such a close match would be very challenging, since the large numbers of degrees of freedom $n-k \sim$\,16\,000--30\,000 mean that the 99\% interval for the $\chi^{-2}$ statistic is narrow, typically $\left[0.98,1.02\right]$.
\\\\
In principle, we could estimate $\sigma_{model}$ by increasing it and broadening the 99\% interval until the value of $\chi^{-2}$ falls within this interval. This would give a meaningful estimate of the `model error' in a particular run. But this would not change the ordering of different model runs, or indeed the final optimal parameters, so we have not included such analysis here.

\subsection{Results} \label{sect23}

\subsubsection{Optimal parameters}\label{optparam}

\begin{figure}
  \resizebox{\hsize}{!}{\includegraphics{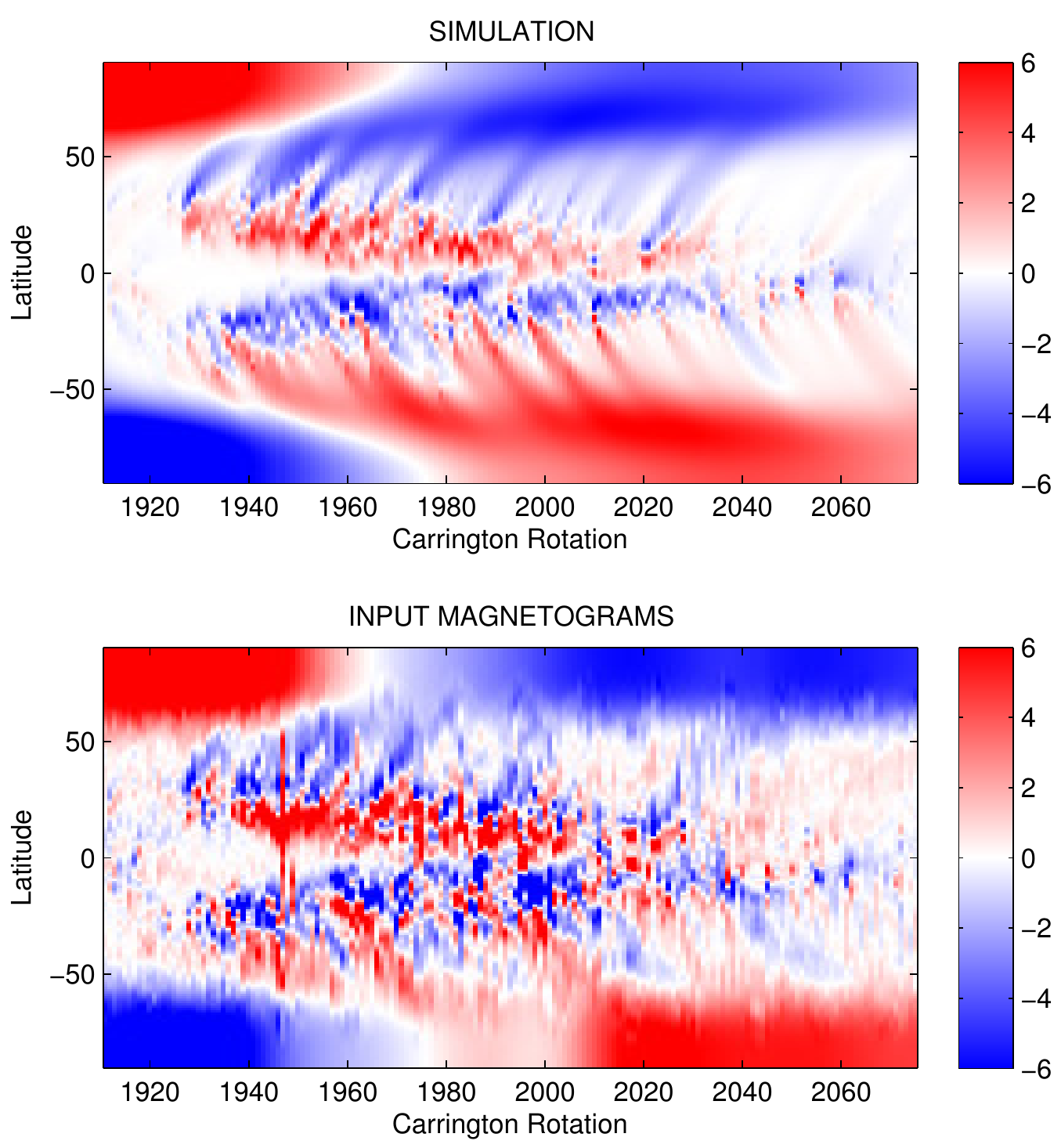}}
  \caption{Top: Butterfly diagram for the optimal parameter 5-set for the 1D model in Table \ref{table1big}(a). Bottom: Ground truth data for Cycle 23.}
  \label{petrie5p}
\end{figure}

In addition to $v_0$, $p$ and $B_0$ as mentioned above, $\eta$ and $\tau$ are also included in the optimization, resulting in 5 parameters to be optimized initially. Maximum and minimum limits are prescribed based on results from literature and observations \citep[e.g.,][]{schrijver02,hathright,yeates14,lemerle}: 
\begin{enumerate}[label=(\roman*)]
  \item $100$\,km$^2$\,s$^{-1} \leq \eta \leq 1500$\,km$^2$\,s$^{-1}$
  \item $5$\,m\,s$^{-1} \leq v_0 \leq 30$\,m\,s$^{-1}$
  \item $0 \leq p \leq 16$
  \item $0$\,yr $\leq \tau \leq 32$\,yr
  \item $0$\,G $\leq B_0 \leq 50$\,G
\end{enumerate}
It should be noted that these ranges are deliberately made wider than results from literature to allow for a deeper exploration into the parameter space and to provide a better understanding of the SFT model. Table \ref{table1big}(a) shows the results of the optimization, with the corresponding optimal butterfly diagram in the top panel of Fig. \ref{petrie5p}, and the interpolated NSO data for Cycle 23 discussed in Sect. \ref{sect22} in the bottom panel for direct qualitative comparison.\\

\begin{table*}
  \caption{Optimal parameter sets for each optimization regime. Entries in bold represent parameters that were fixed for the corresponding run. Upper and lower bounds for acceptable parameter ranges are given in square brackets below each entry. The ranges for regime (a) are presented visually in Fig. \ref{paramrange}.}
  \label{table1big}
  \centering
  \begin{tabular}{c c c c c c c c c}
    \hline
    Regime & $\chi^{-2}$ & $\eta$ & $v_0$ & $p$ & $\tau$ & $B_0$ & $TAF$ & $BPAR$ \T\\
    &  & (km$^2$\,s$^{-1}$) & (m\,s$^{-1}$) & & (yr) & (G) & & (G) \B\\
    \hline
    \hline
    \multicolumn{9}{c}{Cycle 23}\T\B\\
    \hline
    (a) 1D & 0.89 & 351.6 & 14.0 & 3.24 & 2.4 & 16.5 & \textbf{1.00} & n/a \T\\
    & & $\left[229.4,546.9\right]$ & $\left[11.3,22.5\right]$ & $\left[2.98,4.50\right]$ & $\left[1.9,3.5\right]$ & $\left[12.8,20.9\right]$ & & \B\\
    (b) 1D + $TAF$ & 1.09 & 373.5 & 11.0 & 2.44 & 3.7 & 11.7 & 0.55 & n/a \\
    & & $\left[233.3,582.8\right]$ & $\left[8.2,16.5\right]$ & $\left[1.82,2.96\right]$ & $\left[2.9,6.1\right]$ & $\left[8.5,14.8\right]$ & $\left[0.41,0.61\right]$ & \B\\
    (c) 2D + $BPAR$ & 0.65 & 455.6 & 11.2 & 2.76 & n/a & 8.3 & n/a & 39.8 \\
    & & $\left[371.6,651.0\right]$ & $\left[8.6,14.4\right]$ & $\left[1.64,4.71\right]$ & & $\left[5.3,10.2\right]$ & & $\left[31.7,49.4\right]$ \B\\
    (d) 2D + $\tau$ & 0.67 & 453.5 & 9.6 & 2.15 & 4.5 & 12.9 & n/a & \textbf{39.8} \\
    & & $\left[299.5,807.7\right]$ & $\left[6.8,15.2\right]$ & $\left[1.50,3.95\right]$ & $\left[1.6,30.3\right]$ & $\left[6.5,18.4\right]$ & & \B\\
    (e) 1D, fixed $p$ & 0.85 & 361.4 & 8.3 & \textbf{1.87} & 1.9 & 16.3 & \textbf{1.00} & n/a \\
    & & $\left[220.1,642.8\right]$ & $\left[7.4,10.9\right]$ & & $\left[1.5,2.3\right]$ & $\left[11.8,21.4\right]$ & & \B\\
    (f) 2D, fixed $p$ & 0.64 & 482.1 & 11.5 & \textbf{1.87} & n/a & 9.7 & n/a & \textbf{39.8} \\
    & & $\left[356.1,712.9\right]$ & $\left[8.8,15.2\right]$ & & & $\left[7.1,12.8\right]$ & & \B\\
    \hline
    \multicolumn{9}{c}{Cycle 21}\T\B\\
    \hline
    (g) 2D & 0.87 & 455.7 & 9.2 & 2.33 & n/a & 6.6 & n/a & \textbf{39.8} \T\\
    & & $\left[342.7,667.0\right]$ & $\left[6.6,12.0\right]$ & $\left[1.33,3.93\right]$ & & $\left[4.5,9.4\right]$ & & \B\\
    \hline
    \multicolumn{9}{c}{Cycle 22}\T\B\\
    \hline 
    (h) 2D & 0.84 & 506.2 & 8.7 & 2.18 & n/a & 10.5 & n/a & \textbf{39.8} \T\\
    & & $\left[365.1,760.9\right]$ & $\left[6.1,11.7\right]$ & $\left[0.98,3.60\right]$ & & $\left[7.5,13.8\right]$ & & \B\\
    \hline
    \multicolumn{9}{c}{Cycle 24}\T\B\\
    \hline
    (i) 2D & 0.99 & 454.6 & 8.2 & 2.05 & n/a & 4.2 & n/a & \textbf{39.8} \T\\
    & & $\left[292.6,821.7\right]$ & $\left[5.4,12.5\right]$ & $\left[0.62,5.22\right]$ & & $\left[2.6,5.4\right]$ & & \B\\
    \hline
  \end{tabular}
\end{table*}

\noindent The equatorward migration of active regions is well represented by the BMR data, and large poleward surges are reproduced by the model. While the southern polar field reversal is well approximated by the model, the reversal in the northern hemisphere has a delay of approximately 5\,CR. Furthermore, there are multiple weak poleward surges in the simulated butterfly diagram, most noticeably around 2000--2020\,CR, which do not appear in the real butterfly diagram. This is likely to be a by-product of approximating regions as BMRs and overestimating the contribution of flux from smaller regions. This build-up of flux results in a strong polar field that extends to lower latitudes, requiring a short decay timescale as is found in the optimization.

\subsubsection{Parameter analysis} \label{paramana}

\begin{figure*}
  \resizebox{\hsize}{!}{\includegraphics{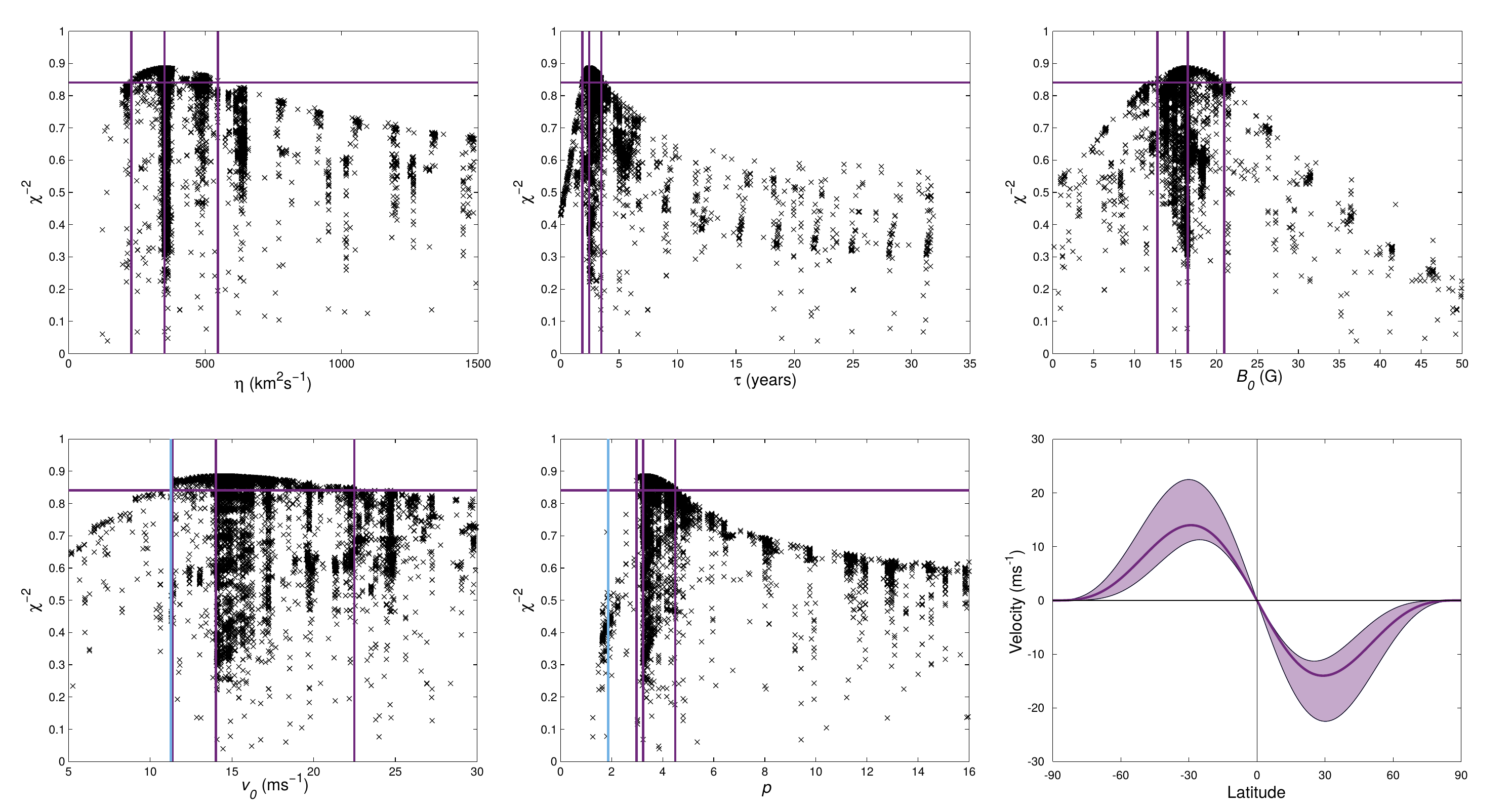}}
  \caption{Scatter plots of every population member for each parameter. The horizontal purple line denotes 95\% of the maximum $\chi^{-2}$. The central vertical purple line is the optimum value for each parameter, with error bars given by the neighbouring vertical purple lines. The vertical blue lines in the bottom left and middle panels are the values obtained from fitting the velocity profile in Equation \ref{meridfloweqn} to observational data (see Sect. \ref{sect4}). The bottom right panel shows the optimal meridional flow profile (bold purple) with acceptable profiles represented by the surrounding purple shading.}
  \label{paramrange}
\end{figure*}

During an optimization run, every single population member generated by \texttt{PIKAIA} can be recorded, and so a range of `acceptable' values can be obtained for each parameter. These can be found in square brackets below each optimal value in Table \ref{table1big}. The upper and lower bounds are taken to be the largest and smallest values for each parameter which produce fits above 95\% of the maximum $\chi^{-2}$. Anything within these limits is classed as `acceptable', though it must be noted that choosing to fix one parameter can alter the optimal solutions and bounds for others. Figure \ref{paramrange} shows such bounds, denoted by the left and right vertical purple lines on each plot, for all parameter populations from the optimization run that produced the optimal set in Table \ref{table1big}(a). The optimal values are highlighted by the central vertical purple lines. Using the limits for $v_0$ and $p$, acceptable meridional flow profiles were also found which are represented by the purple shading in the bottom right panel. The bold purple profile represents the optimum profile.\\

\begin{figure}
  \resizebox{\hsize}{!}{\includegraphics{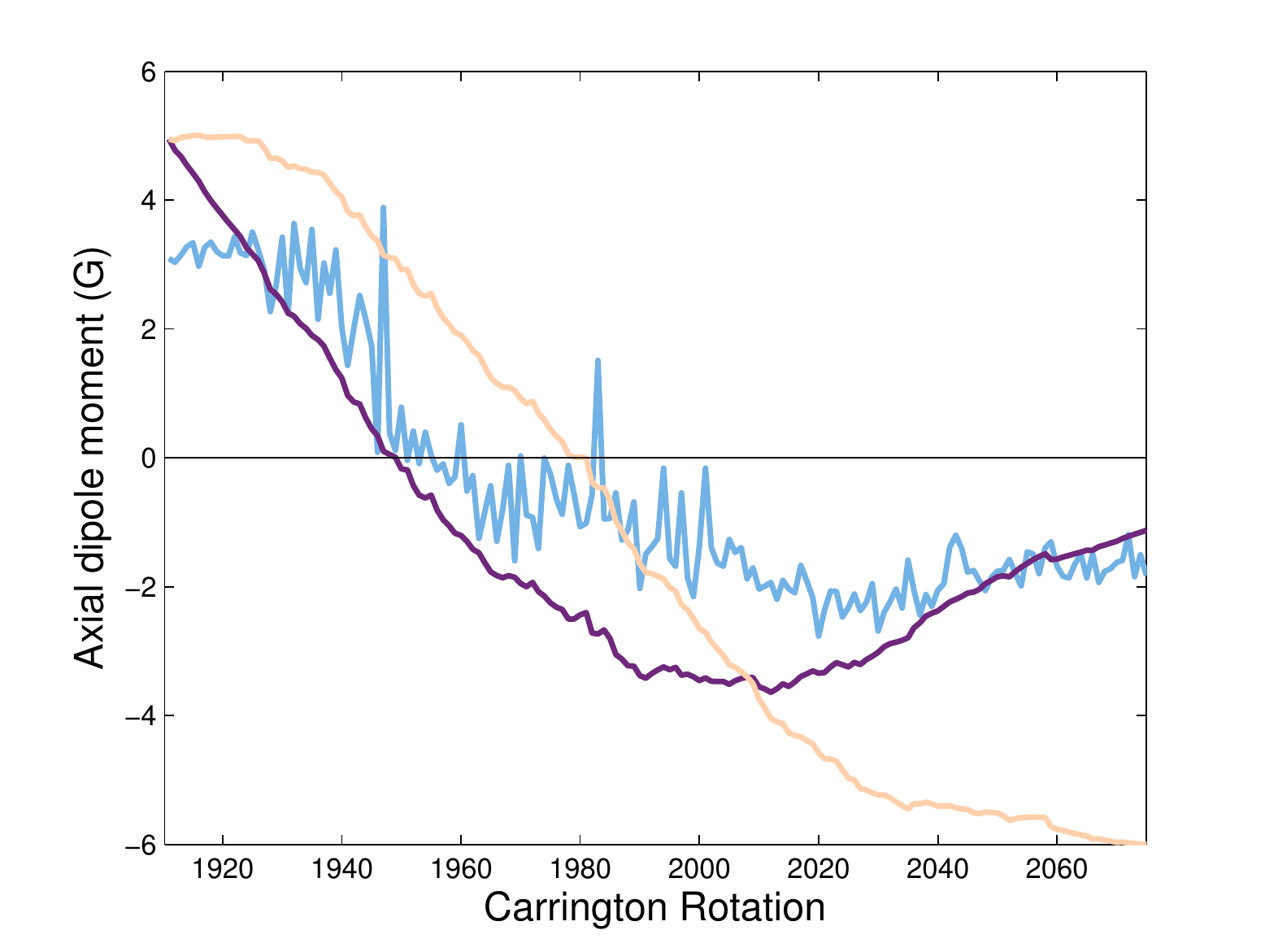}}
  \caption{\bold{Axial dipole moments calculated from observed data (blue), the parameter set in Table \ref{table1big}(a) (purple), and the same parameter set but with the decay term omitted (peach).}}
  \label{axd1d}
\end{figure}

\noindent The diffusion parameter $\eta$ has not yet been accurately measured, though some indirect measurements by \citet{mosher77} and \citet{komm95} have found values within the range of 100--300\,km$^2$\,s$^{-1}$. Early simulations by \citet{leighton64} used values up to 1000\,km$^2$\,s$^{-1}$, though studies by \citet{baumann04}, \citet{wangetal89b} and \citet{wangsh91} decreased it to $\sim$\,600\,km$^2$\,s$^{-1}$, before \citet{wangetal02b} reduced it further to 500\,km$^2$\,s$^{-1}$. Our optimal value of 351.6\,km$^2$\,s$^{-1}$, however, is in better agreement with \citet{yeates14}, who found that $\eta \in \left[200,450\right]$\,km$^2$\,s$^{-1}$ produced a reasonable correlation between the butterfly diagrams, and \citet{lemerle} who found an optimal value of 350\,km$^2$\,s$^{-1}$ within an acceptable range of 240--660\,km$^2$\,s$^{-1}$ for Cycle 21. Furthermore, \citet{schrijver01} and \citet{thibault14} found diffusion coefficients of 300\,km$^2$\,s$^{-1}$ and 416\,km$^2$\,s$^{-1}$ respectively for random-walk-based models, and \citet{cameronetal16} recently used a diffusion of 250\,km$^2$\,s$^{-1}$. The acceptable range in Table \ref{table1big}(a) is broad but can be attributed to multiple degrees of freedom in the optimization. The range covers most values discussed above.
\\\\
The large-scale meridional flow is poorly constrained by observations, as discussed in Sect. \ref{sect21}. Nevertheless, our optimal value of $v_0 = 14$\,m\,s$^{-1}$ is in accordance with both the observations and simulations. Doppler measurements by \citet{ulrich10} estimated the maximum velocity to be between 14--16\,m\,s$^{-1}$ for Cycles 22 and 23. \citet{hathright} obtained an average maximum velocity of 10--12\,m\,s$^{-1}$ for Cycle 23 via magnetic feature tracking, though crucially they observed that the flow is slower (approximately 8\,m\,s$^{-1}$) at cycle maximum and faster (11.5--13\,m\,s$^{-1}$) at minimum. This time-dependence could be added to the model for greater realism, though it is not immediately clear how it could be parametrized in the optimization. Furthermore, they note that many SFT models use meridional flows which go to zero poleward of $\pm 75$\textdegree{} latitude which is not necessarily what is observed, as well as other deviations from observations. \citet{hathupton} prescribed a profile with a maximum velocity of 12\,m\,s$^{-1}$ and \citet{baumann04} used 11\,m\,s$^{-1}$. \citet{yeates14} discovered that a range of 11--15\,m\,s$^{-1}$ improves butterfly diagram correlation, and \citet{wangetal89b} and \citet{wangsh91} found that a range of 7--13\,m\,s$^{-1}$ was acceptable. \citet{wangetal02b} found that a maximum velocity of 20--25\,m\,s$^{-1}$ accurately reproduced solar cycle features, although they used a profile which differs significantly from observations.
\\\\
The theta component of the flexible meridional flow profile in Equation \ref{meridfloweqn} was also used by \citet{mj09}. They obtained a value of $p=2$ by taking an average of helioseismic data weighted by density and fitting it to the sinusoidal profile. In this case $p=2$ does not quite fall into the narrow acceptable range for $p$. The bottom right panel of Fig. \ref{paramrange} shows that values within this range generally correspond to a peak velocity at $\pm 30$\textdegree{} before slowing down to 0\,m\,s$^{-1}$ at $\pm 75$\textdegree{}. As discussed above, this is not necessarily in line with observations. Taking every member of the population above the $95\% \chi^{-2}_{max}$ threshold, we find that the Pearson's correlation coefficient between the acceptable values for $v_0$ and $p$ is $r = 0.86$, indicating that increasing the maximum velocity of the meridional profile generally requires an increase in $p$. A faster velocity means that active regions are transported away from the equator quicker. To counteract this, a larger value of $p$ narrows the band of latitudes at which the velocity is fast, and additionally brings the maximum velocity closer to the equator.
\\\\
Another interesting result is that of 2.4\,yr for the exponential decay time $\tau$. \citet{schrijver02} found that a decay time of 5--10\,yr was necessary to replicate regular polar field reversal, and \citet{yeates14} found that a decay time of 10\,yr produced a better fit between real and simulated butterfly diagrams. \citet{lemerle} found that exponential decay did not have a large effect on the polar field reversal and decided to set $\tau = 32$\,yr, effectively removing the decay term from the model. However, our optimal value for $\tau$ is close to the lower prescribed limit. This could be because of the model trying to account for the unusually weak polar field at the end of Cycle 23 while, for example, \citet{lemerle} performed the optimization for Cycle 21.
\\\\
\bold{Figure \ref{axd1d} highlights the need for the decay term in the 1D model when modelling Cycle 23. The purple curve, which represents the axial dipole moment obtained using the optimal parameter set in Table \ref{table1big}(a), provides a better fit to the observed axial dipole moment (blue) than the peach curve, which is produced from the same parameter set but with the decay term omitted. In this case the polar field becomes too strong and is not weakened enough without the additional decay term}. \bold{\citet{jiang15} found that the decay term was not required to obtain a close match between observed and simulated axial dipole moments, when using active region data from \citet{liulrich2012}. As well as using different active region data, a reduction in tilt angles and a smaller value of $\eta = 250$\,km$^2$\,s$^{-1}$ were included to account for the lack of the decay term. If we use similar parameters for the 1D model, a better axial dipole moment fit is obtained at the expense of an accurate butterfly diagram. Hence we stress that the optimal values in Table \ref{table1big} are with respect to the measure of choice in Equation \ref{fitness}, and other choices of metric might give different results}.
\\\\
Of course, the choice of decay term is not independent of the other parameters, and the Pearson's correlation coefficient between the acceptable values of $v_0$ and $\tau$ is $r = 0.81$: an increase in the flow speed corresponds to less trailing flux being transported to the poles, so a fast decay to weaken the polar field would not be required in the presence of a faster flow.\\

\noindent It should be noted that the decay term in Equation \ref{sfteqn1d} is not directly observed and was added by \citet{schrijver02} to produce regular polar field reversals over multiple cycles. \citet{wangetal02} overcame this problem by increasing the meridional flow speed for stronger cycles. \citet{baumann06} gave a physical explanation of the decay term; namely, it is the effect of radial (i.e., inward) diffusion of flux into the solar interior, which cannot be accounted for directly in the SFT model. In spherical harmonics, different modes decay at different rates, whereas in the exponential decay used by \citet{schrijver02}, all modes decay at the same rate. \citet{baumann06} found that the lowest-order mode decayed the slowest at a rate of 5\,yr (with a corresponding volume diffusion of $\eta = 100$\,km$^2$\,s$^{-1}$), in good agreement with the findings of \citet{schrijver02}. When we include this more sophisticated form of radial diffusion in our model and perform the optimization, we find the lowest-order mode to have an optimal decay time of $\tau_1 = 2.7$\,yr (with a corresponding volume diffusion of $\eta = 190$\,km$^2$\,s$^{-1}$), in good agreement with the decay time found in Table \ref{table1big}(a). Because of this good agreement, we opt to continue to use the original exponential decay parameter.
\\\\
The optimal value for $B_0$ is significantly higher than that used to approximate the initial profile in Fig. \ref{init}. This might be attributed to the choice of functional form in Equation \ref{initeqn}; not enough flux is prescribed between $\pm 45$\textdegree{} and $\pm 80$\textdegree{}, so the algorithm compensates for this by increasing the maximum flux at $\pm 90$\textdegree{}. Alternatively, a strong initial polar field is also required to counteract the short decay time needed to reproduce the weak polar field at the end of Cycle 23.

\subsection{Tilt angles}

\begin{figure}
  \resizebox{\hsize}{!}{\includegraphics{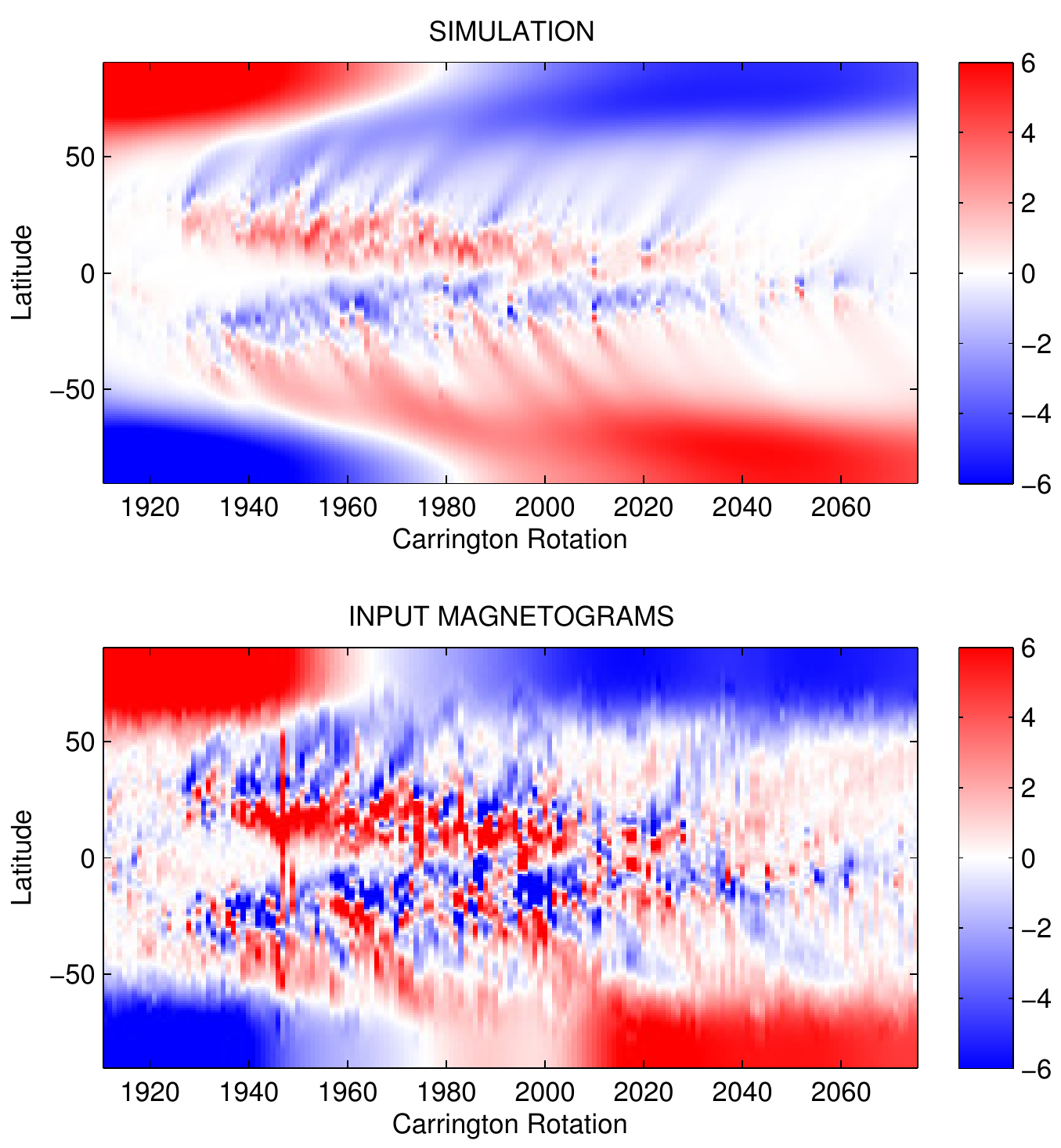}}
  \caption{Top: Butterfly diagram for the optimal parameter 6-set for the 1D model with reduced tilt angles in Table \ref{table1big}(b). Bottom: Ground truth data for Cycle 23.}
  \label{petrie6p}
\end{figure}

Some studies \citep[e.g.,][]{yeates14,jiang11} found that multiplying the tilt angle of each BMR by a scaling factor reduces the polar field strength and improves polar field reversal, since the reduced tilt inhibits equatorial cross-cancellation and hence each magnetic region will contribute less to the axial dipole. A multiplicative tilt angle factor ($TAF$) is included here as an extra parameter to be optimized within the range $0 \leq TAF \leq 1.5$. Table \ref{table1big}(b) shows the results for the 6 parameter case, with the corresponding butterfly diagram in the top panel of Fig. \ref{petrie6p}.
\\\\
The optimal value of 0.55 for $TAF$ is lower than that found by \citet{yeates14} ($TAF \sim 0.8$) and \citet{jiang11} ($TAF \sim 0.72$). It predictably produces a weaker polar field than in the case above where it wasn't included. Given that the main aim of the algorithm is to reduce differences between the real and simulated butterfly diagram pixels, it is reasonable to expect that the optimization algorithm will rely heavily on diffusion and high amplitudes of meridional flow to achieve weak polar fields, although it should be noted that this effect is reduced by the weighting in $\sigma$. Introducing the tilt angle factor as a means of reducing the polar field allows for the decay time to increase and the maximum meridional flow velocity to decrease, suggesting a delicate balance between the parameters and the roles they play in the model. While the polar field strength is better approximated in this case, the active regions are much weaker than in the 5-parameter case, and polar field reversal occurs later in the simulation in both hemispheres. The fitness value of $\chi^{-2} = 1.09$ is above the 99\% interval given in Sect. \ref{pikaia}, which seems to indicate that the model matches the observations better than a randomly chosen map from the observed distribution. This is plainly a limitation of the $\chi^{-2}$ statistic; in particular, it likely indicates the presence of a significant $\sigma_{model}$ term possessing a more complex structure over $\theta$ and $t$. In principle, it could be caused by too large a prescribed $\sigma_{obs}$, or by the relatively strong assumption of independence, or possibly over-fitting of the model, although the latter is unlikely given the small number of parameters in the model. It should be noted that the scaling of tilt angles is not a physical phenomenon, rather a method of reducing the flux in the model, though \citet{cameron10} argued that scaling the tilt angle by a factor of 0.7 mimics the effect of inflows around active regions. Moreover \citet{dasiespuig} found an inverse correlation between cycle strength and tilt angle, suggesting that tilt angle variation plays a significant role in polar field variation.

\section{Two-dimensional model} \label{sect3}

\begin{figure}
  \resizebox{\hsize}{!}{\includegraphics{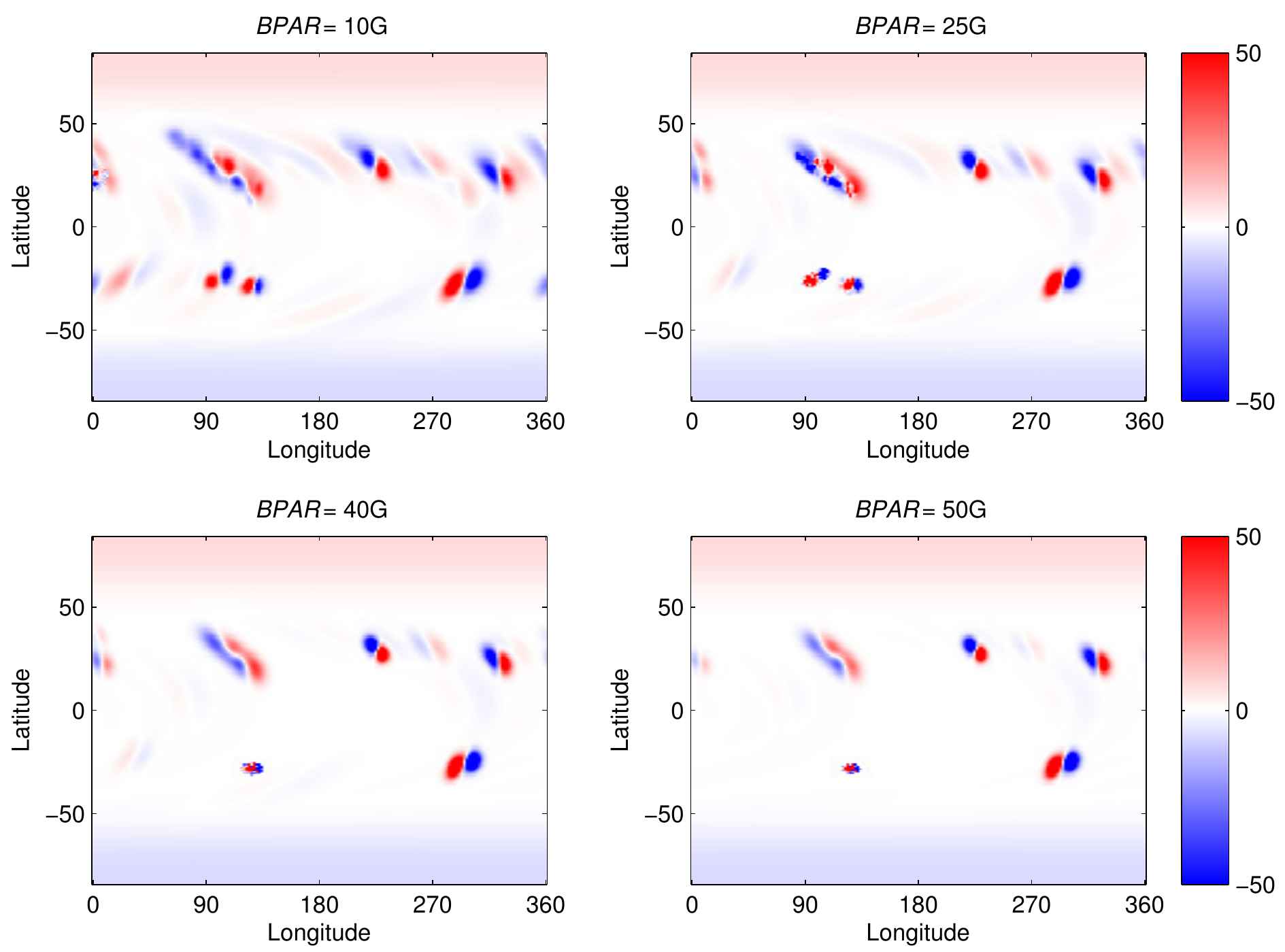}}
  \caption{Four snapshots of 1928\,CR from simulations with active regions selected by different values of the magnetic flux threshold $BPAR$ and all other parameters fixed. Here $BPAR$ increases from left to right and top to bottom.}
  \label{bpars}
\end{figure}

\citet{yeates15} developed a 2D model\footnotemark\footnotetext{https://github.com/antyeates1983/sft\_data} which assimilates specific shapes of magnetic regions into the simulation on the day of emergence. The aim of the model is to better assimilate strong, multipolar regions, which are not accurately portrayed in a simpler bipolar form, as in the 1D model above, with the hope of simulating a more realistic photospheric field. This selection feature requires the model to be 2D. The model is fully automated, providing consistent highlighting of strong magnetic regions, and is designed to replace pre-existing regions rather than superimposing new ones. The SFT equation for the radial component of the magnetic field in 2D, $B\left(\theta ,\phi ,t\right)$, is:
\begin{align}
  \frac{\partial{B}}{\partial{t}} =& - \omega \left(\theta\right)\frac{\partial{B}}{\partial{\phi}} + \frac{\eta}{R_{\odot}^2}\left[\frac{1}{\sin\theta}\frac{\partial}{\partial{\theta}}\left(\sin\theta\frac{\partial{B}}{\partial{\theta}}\right) + \frac{1}{\sin^2\theta}\frac{\partial ^2{B}}{\partial{\phi ^2}}\right]\nonumber\\
  &-  \frac{1}{R_{\odot}\sin\theta}\frac{\partial}{\partial{\theta}}\Big(v\left(\theta\right)\sin\theta\,B\Big) - \frac{1}{\tau}B + S(\theta,\phi,t) ,
\end{align}
where $\omega$ represents differential rotation and all other parameters are identical to those in Equation \ref{sfteqn1d}. Note that differential rotation averaged out and so played no role in the axisymmetric 1D model. Rather than using a spectral method like the 1D model, the evolution equations (for the vector potential) are solved in the Carrington frame using a finite-difference method on a spatial grid of 180 cells equally spaced in sine-latitude and 180 cells equally spaced in longitude. Unlike meridional flow, differential rotation is well constrained by observations, and in the model is parametrized as \citep{snodgrass90}:
\begin{equation}
  \omega \left(\theta\right) = 0.521 - 2.396\cos^2\theta - 1.787\cos^4\theta\,\mbox{deg}\,\mbox{day}^{-1} .
\end{equation}
The 2D model contains a parameter $BPAR$ which determines the threshold above which magnetic flux is assimilated into the simulation, in the form of individual strong-flux regions. \citet{yeates15} chose the threshold of $BPAR = 15$\,G in order that the difference between the observed unsigned flux and simulated unsigned flux (due to the smoother magnetic field distribution) remained approximately constant. This parameter is subsequently added to the optimization. If given enough freedom, the algorithm would gradually reduce $BPAR$, allowing more and more magnetic regions to be inserted until the original synoptic map is essentially copied in (analogous to $BPAR \sim 0$\,G). To avoid this, the lower bound is set at $10$\,G with an upper bound of $50$\,G. Figure \ref{bpars} shows snapshots of 1928\,CR from four simulations with alternative values of $BPAR$ between 10\,G and 50\,G, and all other parameters fixed. As the threshold $BPAR$ increases, fewer active regions are assimilated into the simulation.

\subsection{Five-parameter optimization}

\begin{figure}
  \resizebox{\hsize}{!}{\includegraphics{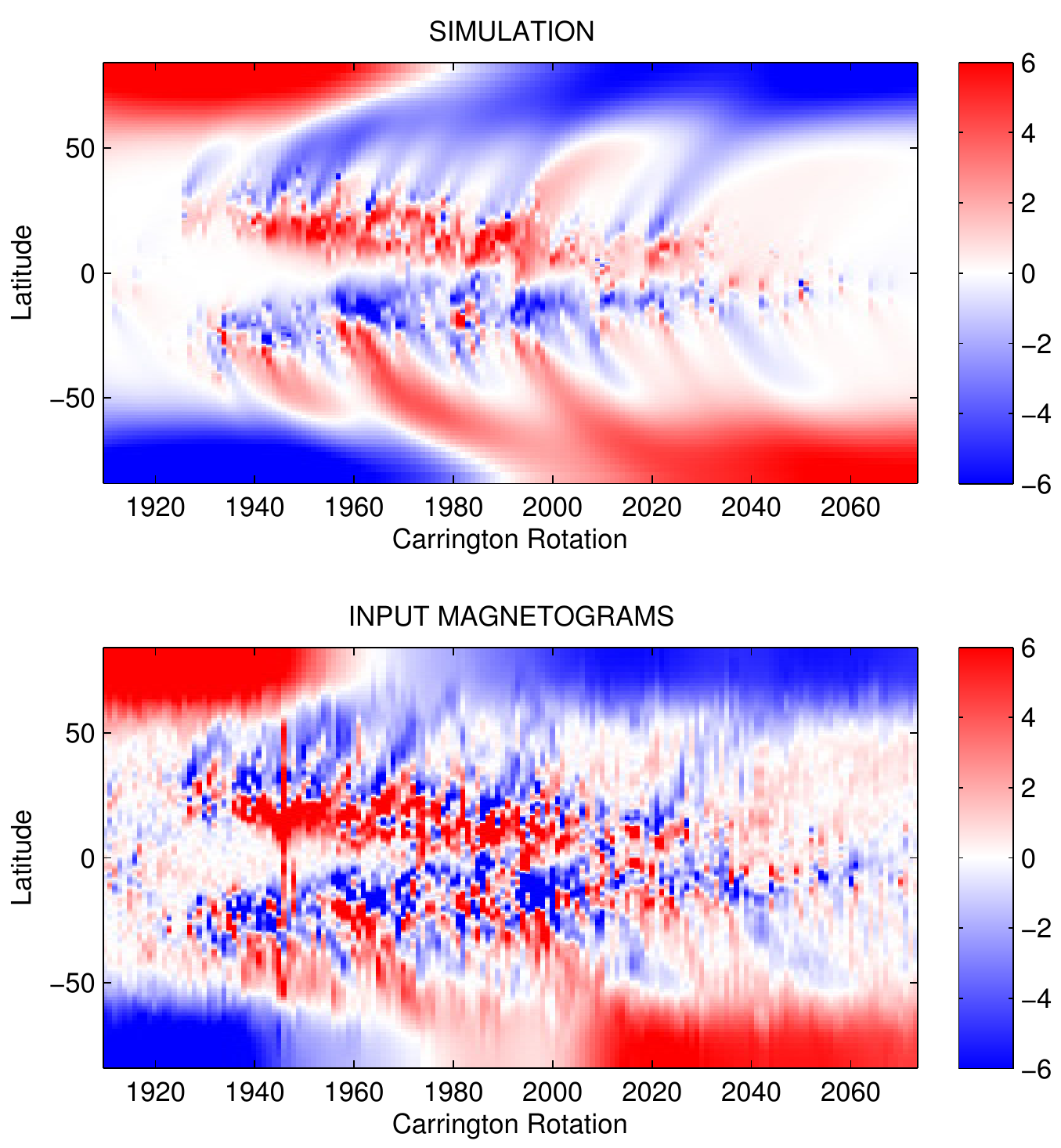}}
  \caption{Top: Butterfly diagram for the optimal parameter 5-set for the 2D model with varying $BPAR$ in Table \ref{table1big}(c). Bottom: Ground truth data for Cycle 23.}
  \label{bpar40compare}
\end{figure}

\begin{figure}
  \resizebox{\hsize}{!}{\includegraphics{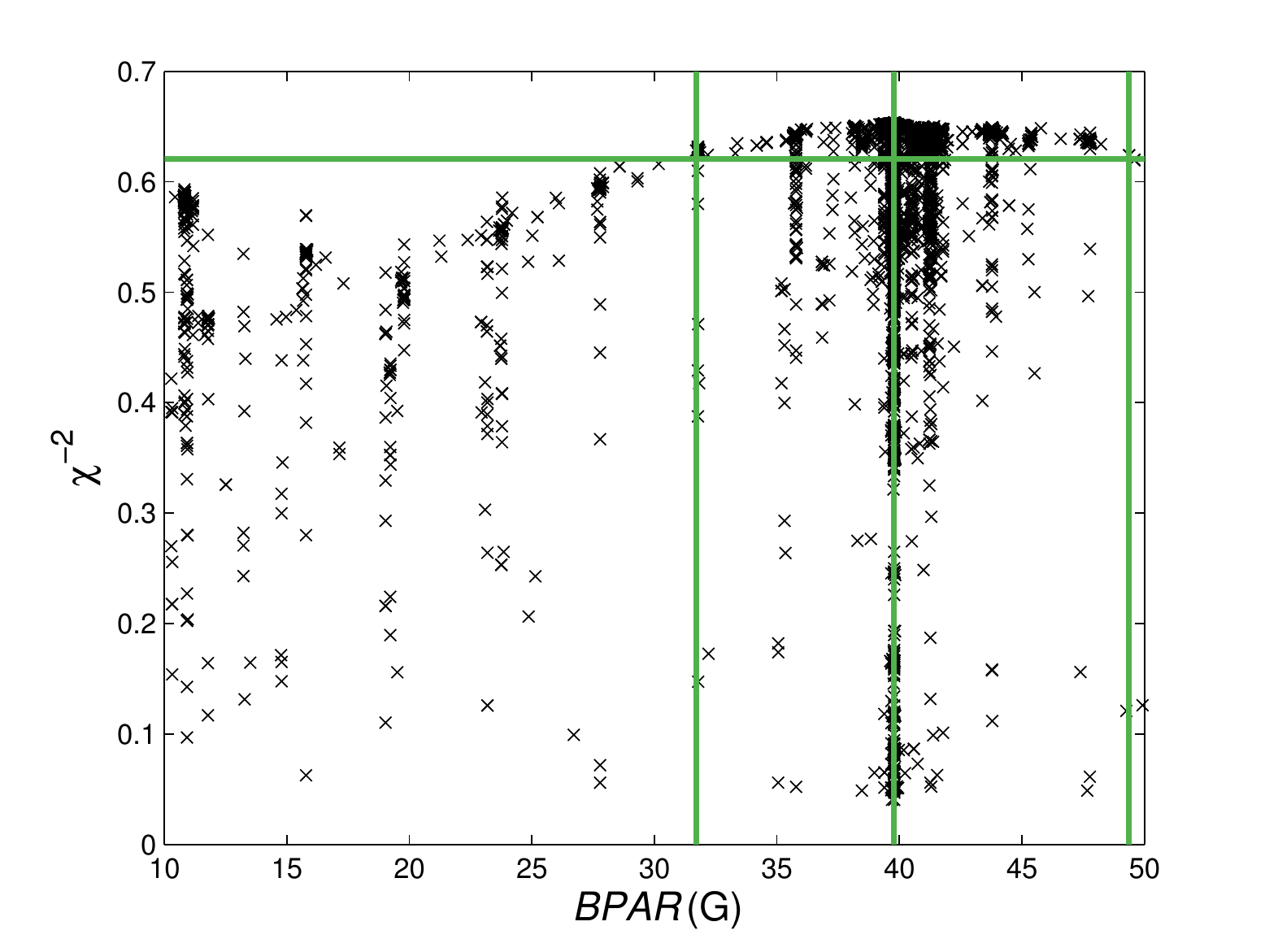}}
  \caption{Each population member for the 5-parameter optimization of the 2D model with $BPAR$ restricted to $BPAR \geq 10$\,G. The horizontal green line denotes 95\% of the maximum $\chi^{-2}$. The central vertical green line is the optimum value for each parameter, with error bars given by the neighbouring vertical green lines.}
  \label{bpar40bpar}
\end{figure}

The synoptic magnetograms from NSO Kitt Peak are used to identify strong regions for assimilation. For simplicity, \citet{yeates15} did not incorporate exponential decay into the model as in Equation \ref{sfteqn1d}. We perform optimization runs for the model both without decay and with the decay term included. Initially we consider the former case. Aside from $BPAR$, parameters are given the same upper and lower limits as in Sect. \ref{optparam}. Table \ref{table1big}(c) shows the results of the optimization. The corresponding butterfly diagram is shown in the top panel of Fig. \ref{bpar40compare}.
\\\\
The 2D model qualitatively improves the butterfly diagram, with active regions predictably more accurate, leading to the inclusion of more poleward surges in the simulation which can be identified in the observed butterfly diagram (though the gradient and strength of each surge is not always correct), and a more realistic polar field. The optimal parameters in Table \ref{table1big}(c) are within the range of other results from simulations and observations described in Sect. \ref{paramana}. A diffusivity of $\eta = 455.6$\,km$^2$\,s$^{-1}$ is a stronger diffusivity than in the 1D model, but the inclusion of an exponential decay term is expected to reduce this. An increased diffusivity is somewhat supported by \citet{virtanen}, who used a value of $\eta = 400$\,km$^2$\,s$^{-1}$ in the same 2D model but for a single simulation of multiple cycles. The range and optimal value for $v_0$ is lower than for the original 1D case, indicating that there can be inherent differences between models. Moreover, \citet{virtanen} found that a value of $v_0 = 11$\,m\,s$^{-1}$ correctly reproduced shapes of poleward surges and polar fields, in excellent agreement with our optimal value.
\\\\
Figure \ref{bpar40bpar} shows every generated value of $BPAR$ against $\chi ^{-2}$. The central vertical line indicates the optimum value of $39.8$\,G, with the left and right vertical lines denoting the acceptable range for $BPAR$, as in Fig. \ref{paramrange}. The value of $15$\,G used by \citet{yeates15} is outside of this range, and for the remainder of the 2D optimizations, $BPAR$ is fixed at the optimal value of $39.8$\,G to attain consistency. This should ensure that only newly emerging regions are inserted for each Carrington rotation. However, the presence of the strong mid-latitudinal region of positive flux in the northern hemisphere at 2000--2020\,CR could be attributed to the choice of large $BPAR$, since smaller regions of negative flux which would otherwise cancel out this positive flux are not being assimilated. The bottom left panel of Fig. \ref{bpars} closely represents the scenario when $BPAR$ is set at its optimal value. \citet{virtanen} used a threshold of $BPAR = 50$\,G, and this lies just outside of our acceptable range. Comparing the bottom two panels of Fig. \ref{bpars}, however, shows that the differences between our optimal value and their chosen value are minor.

\subsection{Incorporating exponential decay}

\begin{figure}
  \resizebox{\hsize}{!}{\includegraphics{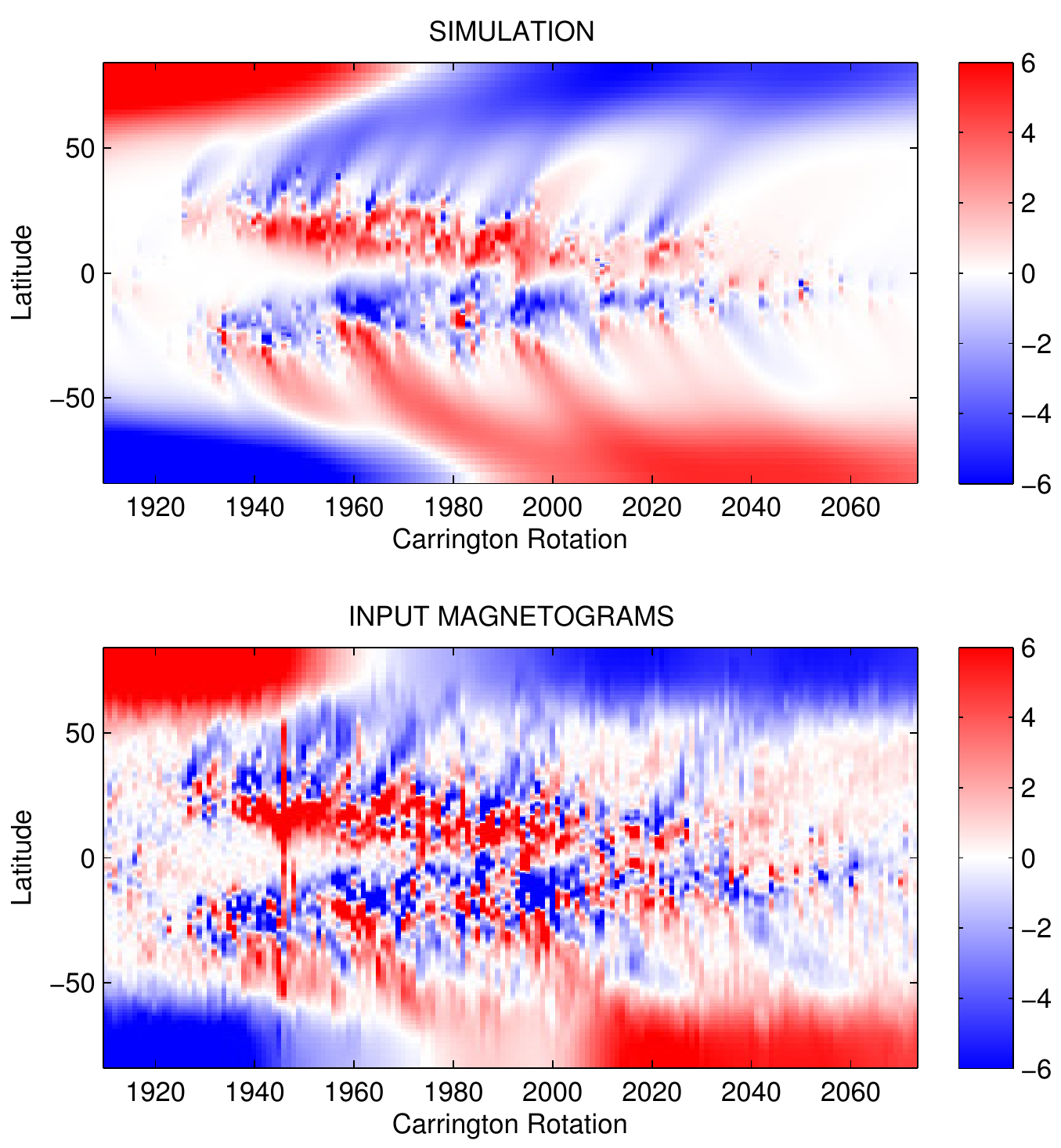}}
  \caption{Top: Butterfly diagram for the optimal parameter 5-set for the 2D model with varying $\tau$ in Table \ref{table1big}(d). Bottom: Ground truth data for Cycle 23.}
  \label{2dtau}
\end{figure}

\begin{figure}
  \resizebox{\hsize}{!}{\includegraphics{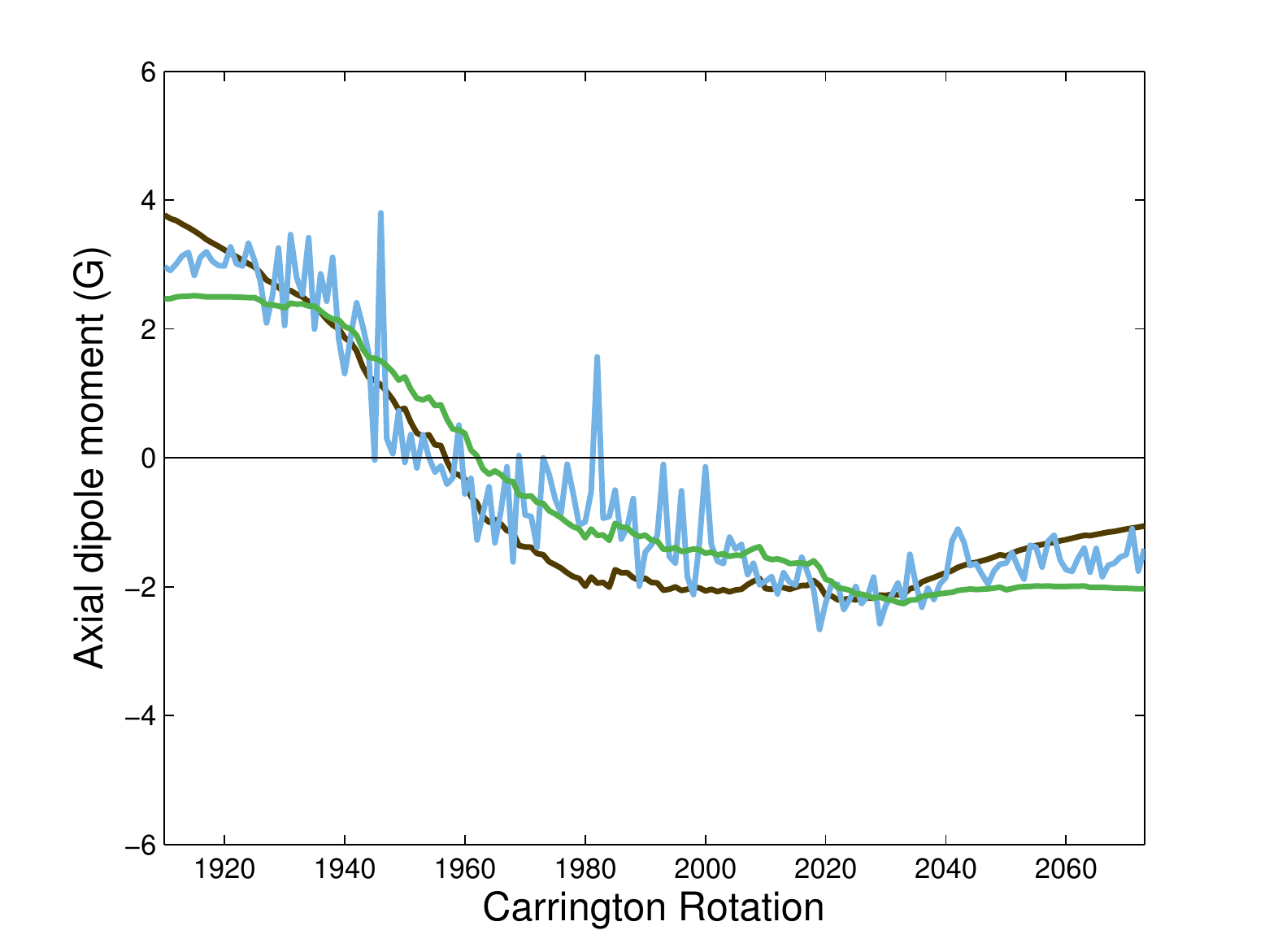}}
  \caption{\bold{Axial dipole moments calculated from observed data (blue), the parameter set in Table \ref{table1big}(c) (green), and the parameter set in Table \ref{table1big}(d) (brown).}}
  \label{axd2d}
\end{figure}

As discussed above, the decay parameter $\tau$ was originally added to the SFT model to produce regular polar field reversals. The 2D model did not initially take account of this decay time, but we incorporate it to assess whether the optimal value in Table \ref{table1big}(a) is reasonable.
\\\\
As shown in Fig. \ref{2dtau}, including the decay term improves timing of polar field reversal by 5--10\,CR, but is not enough to replicate the observed reversal time. Poleward surges are generally wider in the simulation, leading to the reduction of some mid-latitude features, most notably the strong surge of positive flux at 2000--2020\,CR in the northern hemisphere, which is more visible in Fig. \ref{bpar40compare}.
\\\\
The optimization results are shown in Table \ref{table1big}(d). Surprisingly, the addition of an extra decay term induces a minimal decline in diffusion, and it is not enough to bring it down to $351.6$\,km$^2$\,s$^{-1}$ as found in the 1D case. Rather, $B_0$ increases to account for the stronger decay of the polar fields in this regime. Most significantly, we obtain an optimal value of $\tau = 4.5$\,yr. This is higher than the optimum found in the 1D model and in closer agreement with \citet{schrijver02}, although the acceptable range is considerably wider towards the upper limit, indicating that a decay term may not be required in the assimilative model. This is supported by the value of $\chi ^{-2}$ which does not increase significantly with the addition of the decay term. \bold{Furthermore, Fig. \ref{axd2d} shows that the axial dipole moments calculated using the optimal parameter sets for the 2D model, with and without the exponential decay term (brown and green curves respectively), both produce good fits to the observed profile (blue). This indicates that the method of new flux assimilation in the 2D model is better able to account for the weak polar field at the Cycle 23/24 minimum than the idealized BMRs used in the 1D model, since it does not require an additional decay term}. Coupled to the short optimal decay timescale are smaller optimal values for $v_0$ and $p$, suggesting that the relationships and correlations discussed in Sect. \ref{sect23} also hold for the 2D case.

\section{Comparison with meridional flow observations} \label{sect4}

\begin{figure}
  \resizebox{\hsize}{!}{\includegraphics{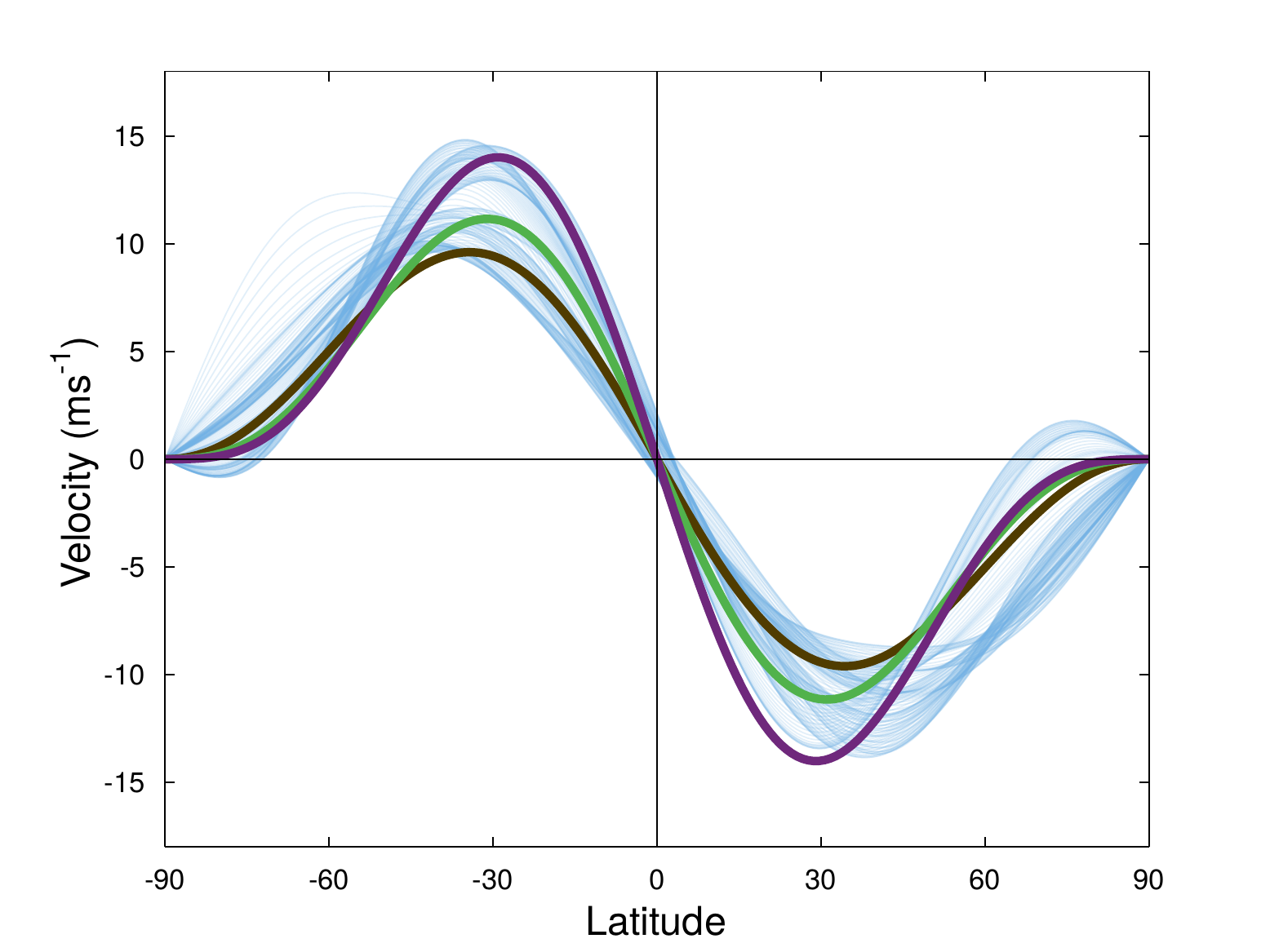}}
  \caption{Comparison of various meridional flow profiles: observed for each CR (blue), 1D optimum (purple), 2D optimum (green) and 2D optimum with decay (brown).}
  \label{meridflows}
\end{figure}

Although observations of the meridional flow are not yet fully reliable, we can use the data that are available to try to add a further constraint to the optimization.
\\\\
David Hathaway kindly provided us with measurements of the meridional flow for Solar Cycle 23, calculated by tracking features in images from the \textit{Solar and Heliospheric Observatory (SOHO)} Michelson Doppler Imager \citep[MDI,][]{mdi}. The data were supplied as coefficients of the following parametrization:
\begin{align}
  v\left(\theta\right) =& \,\Big(C_0 + C_1 \cos\theta + C_2\cos^2 \theta + C_3\cos^3 \theta +C_4\cos^4 \theta\nonumber\\
  &+ C_5\cos^5 \theta\Big) \sin\theta .
\end{align}
The meridional flow measurements for each Carrington rotation are shown in Fig. \ref{meridflows} (blue curves). The observations tend to follow either a fast or slow flow, highlighted by denser blue areas, indicating the dependence on time and that the flow transitions between the two extremes throughout the cycle. Additionally, for a small number of Carrington rotations an equatorward counterflow is observed at high latitudes, though it should be noted that such a counterflow was not visible in HMI data \citep{countercell}. The choice of flexible profile in Equation \ref{meridfloweqn} does not allow for this phenomenon.
\\\\
The optimal profile using the parameters from the 1D optimization in Table \ref{table1big}(a) is shown in purple in Fig. \ref{meridflows} for comparison. Whilst the observed and optimal profiles are similar in shape, the optimal profile is too fast and reaches its peak at a slightly lower latitude. Moreover, the observed profiles tend to extend beyond $\pm 75$\textdegree{} but the optimal profile chooses to go to zero throughout the polar regions, giving a possible explanation as to why many SFT models incorporate this feature. Furthermore, the 1D optimal profile remains almost completely within the bounds given by the observations, excluding at its peak in the northern hemisphere for which asymmetry in the observations can be held responsible.
\\\\
The green and brown profiles in Fig. \ref{meridflows} represent the optima for the 2D model excluding and including exponential decay respectively. Both profiles are fully contained within the observational limits, except for a small section of the brown curve in the southern hemisphere which is due to a lower than average maximum velocity. Of the three optimal profiles, the 2D regime without decay matches the average observed profile the closest, whilst the decay-enhanced flow is slightly slower (though \citet{hathright} observed speeds of 8\,m\,s$^{-1}$ at cycle maximum). It does, however, continue to latitudes poleward of $\pm 70$\textdegree{}, almost emulating the observational data. One limitation of tracking magnetic features to measure the meridional flow is that it is not always easy to distinguish between the effects of the meridional flow and the effects of supergranular diffusion. For this reason, flows derived from feature tracking tend to peak at higher latitudes \citep[e.g.,][Fig. 1]{dikpati10}, giving a possible explanation as to why the observed curves in Fig. \ref{meridflows} tend to peak at higher latitudes than the modelled curves.\\

\begin{figure}
  \resizebox{\hsize}{!}{\includegraphics{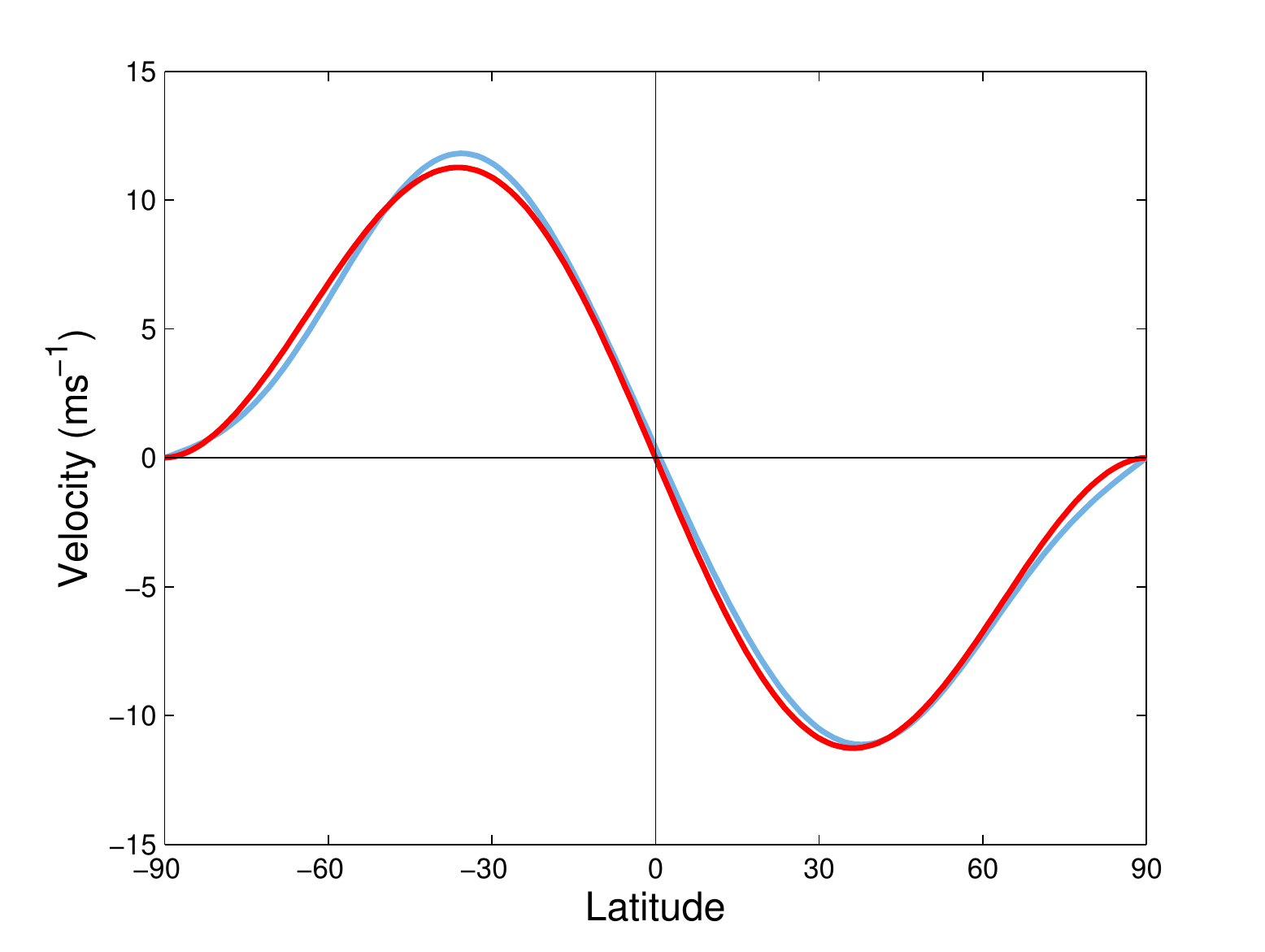}}
  \caption{Comparison of average observed (blue) and fitted (red) meridional flow profiles.}
  \label{meridflowfit}
\end{figure}

\noindent We use a non-linear least-squares fitting method to fit the parametrized form of the meridional flow in Equation \ref{meridfloweqn} to the average observed coefficients given by David Hathaway to ensure it is actually possible to match the observed profile. The average observed and fitted profiles, shown in Fig. \ref{meridflowfit} (blue and red respectively), match closely for $v_0 = 11.3$\,m\,s$^{-1}$ and $p = 1.87$, and slight asymmetry in the average observed profile is confirmed. This value of $p$ is close to that of \citet{mj09} and is within the acceptable ranges for $p$ in the above 2D regimes, but is outside the equivalent range in the 1D optimization run, whence we infer that the 1D model requires the maximum velocity to be closer to the equator than is observed.\\

\begin{figure}
  \resizebox{\hsize}{!}{\includegraphics{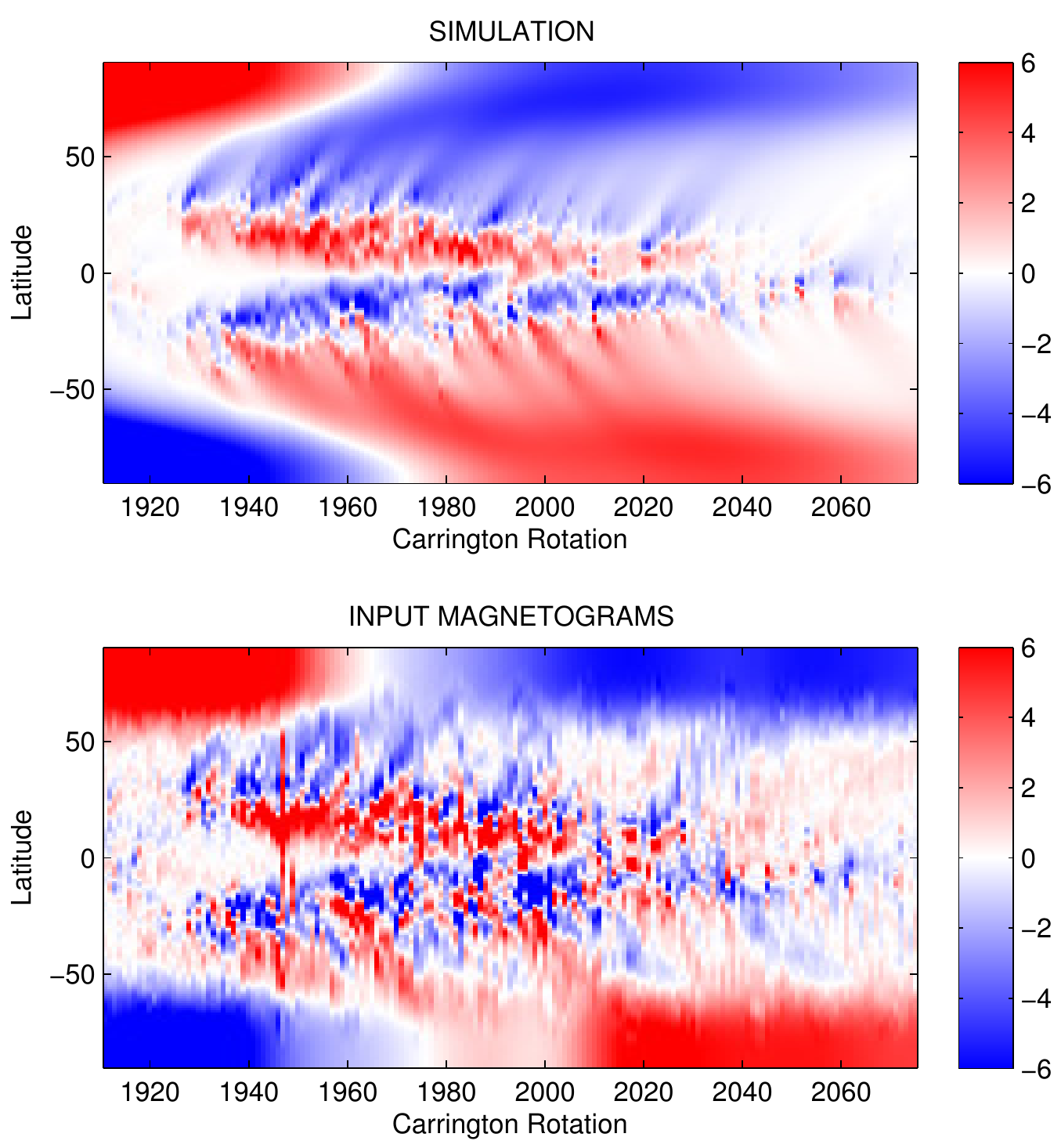}}
  \caption{Top: Butterfly diagram for the optimal parameter 4-set for the 1D model with fixed $p=1.87$ in Table \ref{table1big}(e). Bottom: Ground truth data for Cycle 23.}
  \label{fixedp1d}
\end{figure}

\noindent Given that the parametrization is able to closely fit the observed data, we could fix one of the velocity-related parameters, say $p$, to the observed value and perform optimization runs for the two models. We choose $p$ because the model is generally less sensitive to the choice of $v_0$, and $p = 1.87$ is outside the acceptable range for the 1D model.
\\\\
The optimization results with $p$ fixed in the 1D model are shown in Table \ref{table1big}(e). The value $p = 1.87$ corresponds to a maximum velocity at $\pm 35$\textdegree{}, meaning poleward transport is slower at low latitudes. This results in more flux cancellation across the equator and so more trailing flux is present in the transport regions, as observed in the top panel of Fig. \ref{fixedp1d}. This feature appears to be a common occurrence in the standard SFT model (cf. Figs. \ref{petrie5p} and \ref{petrie6p}). The upshot of this numerically is that the selected decay time of 1.9\,yr is even shorter than in the original 1D case to counteract the large amounts of flux accumulating at the poles. This couples with a slow velocity, made even slower by the small value of $p$, adhering to the relationship found in Sect. \ref{sect23}. The timing of polar field reversal, meanwhile, is reproduced reasonably accurately. Except for a marginally smaller value of $\chi^{-2}$, fixing $p$ does not significantly hinder the quantitative performance of the 1D model, even though $p = 1.87$ is not in the acceptable parameter range for regime (a).\\

\begin{figure}
  \resizebox{\hsize}{!}{\includegraphics{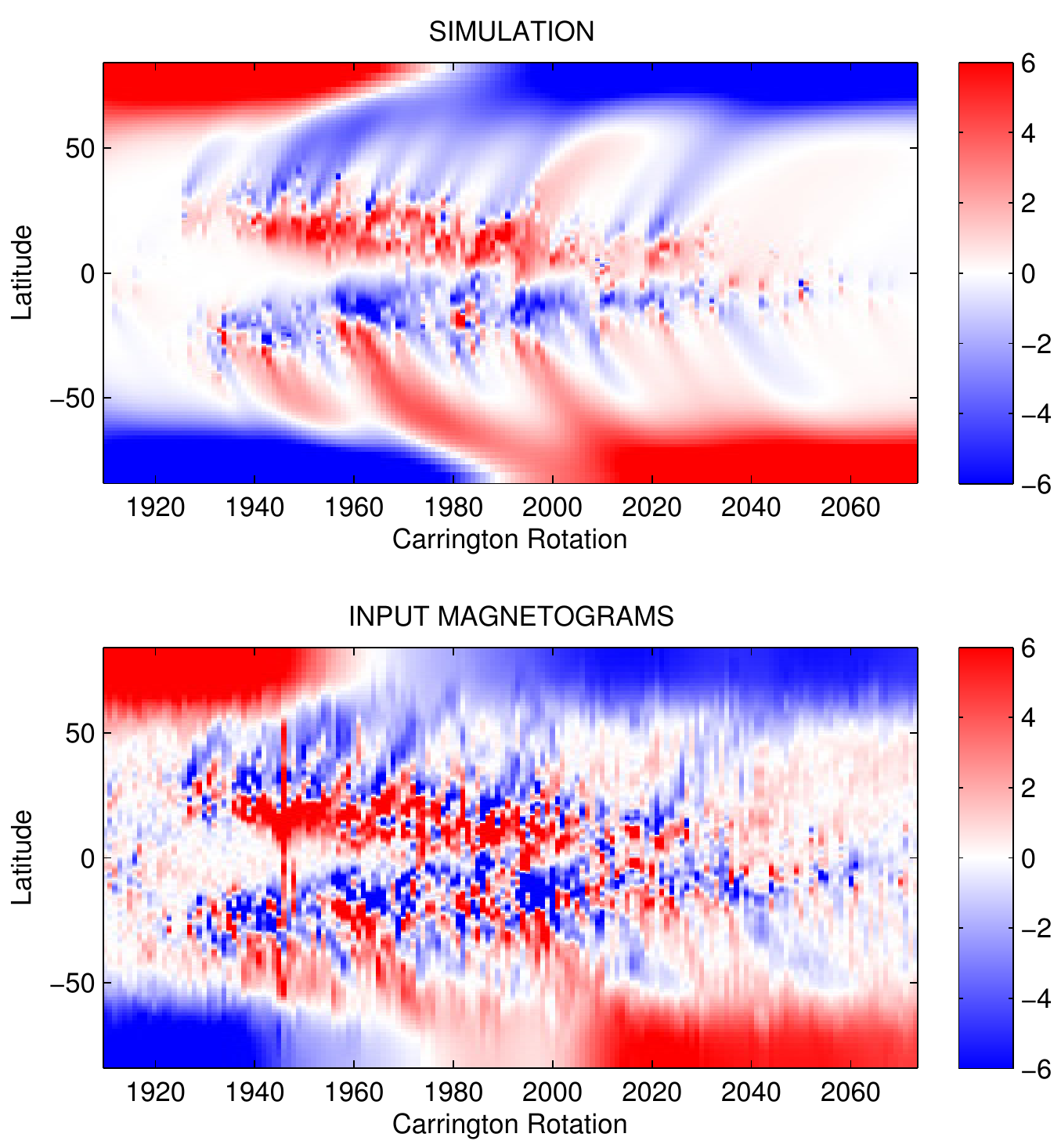}}
  \caption{Top: Butterfly diagram for the optimal parameter 3-set for the 2D model with fixed $p=1.87$ in Table \ref{table1big}(f). Bottom: Ground truth data for Cycle 23.}
  \label{fixedp2d}
\end{figure}

\noindent With the higher-latitudinal velocity peak and the absence of $\tau$ in the 2D model, the resulting diffusion value given in Table \ref{table1big}(f) is slightly larger than in previous regimes. Contrary to expectation, the optimal maximum velocity is higher than the previous 2D cases, but still with wide error bounds. Given that $p = 1.87$ lies within the acceptable range in regime (c), it is reasonable to expect that optimal values and associated ranges would be in line with results in Sect. \ref{sect3} and hence observations and previous studies. Consequently the optimal butterfly diagram (top panel of Fig. \ref{fixedp2d}) confirms this, offering only subtle changes to Fig. \ref{bpar40compare}, for example a polar field restricted to higher latitudes due to the increase in diffusivity.

\section{Other solar cycles} \label{sect5}

With its automated assimilation of active region data, the 2D model can easily be adapted for other cycles, provided there is sufficient data available. Evaluations of Cycles 21, 22 and 24 (up to the end of 2015) using NSO data have been carried out to search for cycle-to-cycle variation.

\subsection{Cycle 21} \label{sectcycle21}

\begin{figure}
  \resizebox{\hsize}{!}{\includegraphics{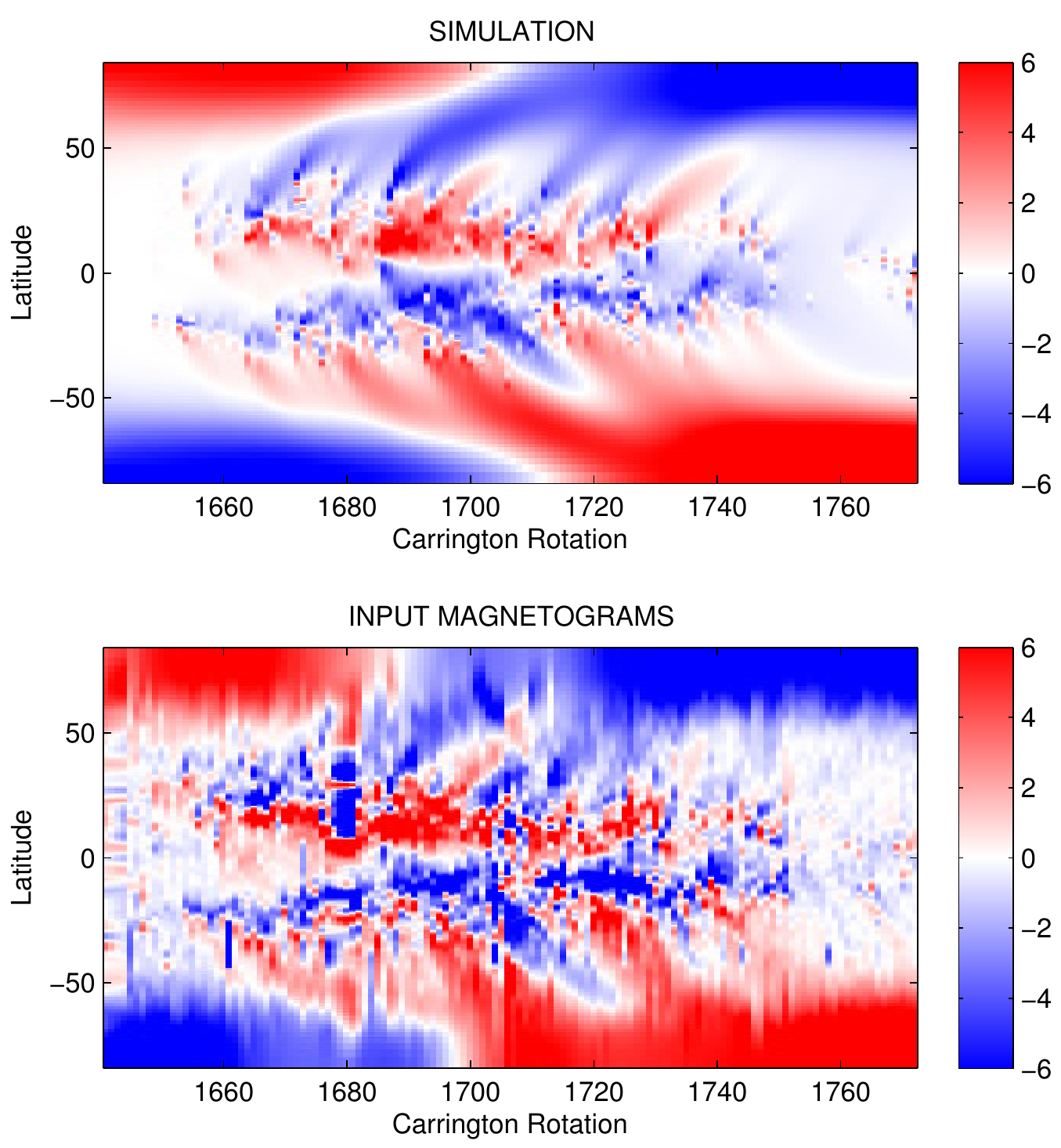}}
  \caption{Top: Butterfly diagram for the optimal parameter 4-set for the 2D model in Table \ref{table1big}(g). Bottom: Ground truth data for Cycle 21.}
  \label{2d4p21}
\end{figure}

Table \ref{table1big}(g) shows the optimum parameters for Cycle 21 (1\ts{st} May 1976--10\ts{th} March 1986). Both $\eta$ and $v_0$ are in agreement with previous studies. Most notably, $v_0 = 9.2$\,m\,s$^{-1}$ is slower than the maximum speed of Cycle 23, supporting \citet{hathupton}: a faster flow in Cycle 23 would have resulted in a weaker polar field at cycle minimum since leading flux is taken away from the equator quickly and so has less time to cancel across the equator. This optimum value, however, is just outside the range of 10--13.2\,m\,s$^{-1}$ as found by \citet{komm93} using feature tracking during Cycle 21. Conversely, this range overlaps with a large portion of the 95\% confidence interval obtained by the optimization population.
\\\\
The interpolated NSO data is shown in the bottom panel of Fig. \ref{2d4p21} with the corresponding simulated butterfly diagram in the top panel of Fig. \ref{2d4p21}. Aside from a negative-polarity observational artefact in the northern hemisphere at 1680\,CR, many features of active regions are well reproduced. There are three instances of large concentrations of opposite flux being transported polewards in the northern hemisphere; the latter of these is over-estimated by the simulation and this could be attributed to the model incorrectly reading in the corresponding emergence region. Polar field reversal for both poles is too late in the model, particularly in the northern hemisphere where the difference is in the region of 10\,CR.
\\\\
\citet{lemerle} performed a similar optimization process for Cycle 21 using a 2D model and a BMR database compiled by \citet{wangsh89}. Although they used a different parametrization for the meridional flow and different sources of flux, their optimal parameter ranges for $\eta$ and $v_0$ were in good agreement with those in Table \ref{table1big}(g). The diffusion coefficient $\eta = 455.7$\,km$^2$\,s$^{-1}$ lies within their acceptable range of 240--660\,km$^2$\,s$^{-1}$ and $v_0 = 9.2$\,m\,s$^{-1}$ falls between 8--18\,m\,s$^{-1}$ as calculated by \texttt{PIKAIA} in their study. They used the following functional form to represent meridional flow:
\begin{equation}\label{lemerleeqn}
  v\left(\theta\right) = -v_0\,\mbox{erf}^q\left(v\sin\theta\right) \mbox{erf}\left(w\cos\theta\right) .
\end{equation}
The optimization returned values of $v_0 = 12$\,m\,s$^{-1}$, $q = 7$, $v = 2$ and $w = 8$. This gave a profile similar to that of \citet{wangetal02b}, but with a less extreme steep gradient at the equator. However, when normalized, the profile shape was comparable to the observed profile formed from Doppler measurements obtained by \citet{ulrich10}, and the observed profile lay well within the error bars for the optimal solution, except for some return flows at high latitudes, which were not incorporable in Equation \ref{lemerleeqn}, mirroring the limitation of our parametrization in Equation \ref{meridfloweqn}. Using a non-linear least-squares fitting method, we are able to attempt to fit the functional form in Equation \ref{meridfloweqn} to the versatile meridional profile in Equation \ref{lemerleeqn}. The best fit corresponds to values of $v_0 = 13.6$\,m\,s$^{-1}$ and $p = 3.88$. This value for $v_0$ is in agreement with observations and acceptable ranges for other regimes, but is above the range for Cycle 21. Despite lying within the acceptable range, $p = 3.88$ favours the high values for $p$ obtained from optimization runs as opposed to the lower values extracted from observational data. This could suggest an inherent flaw within the SFT model whereby the model performs better when the maximum velocity is prescribed to be closer to the equator.

\subsection{Cycle 22}

\begin{figure}
  \resizebox{\hsize}{!}{\includegraphics{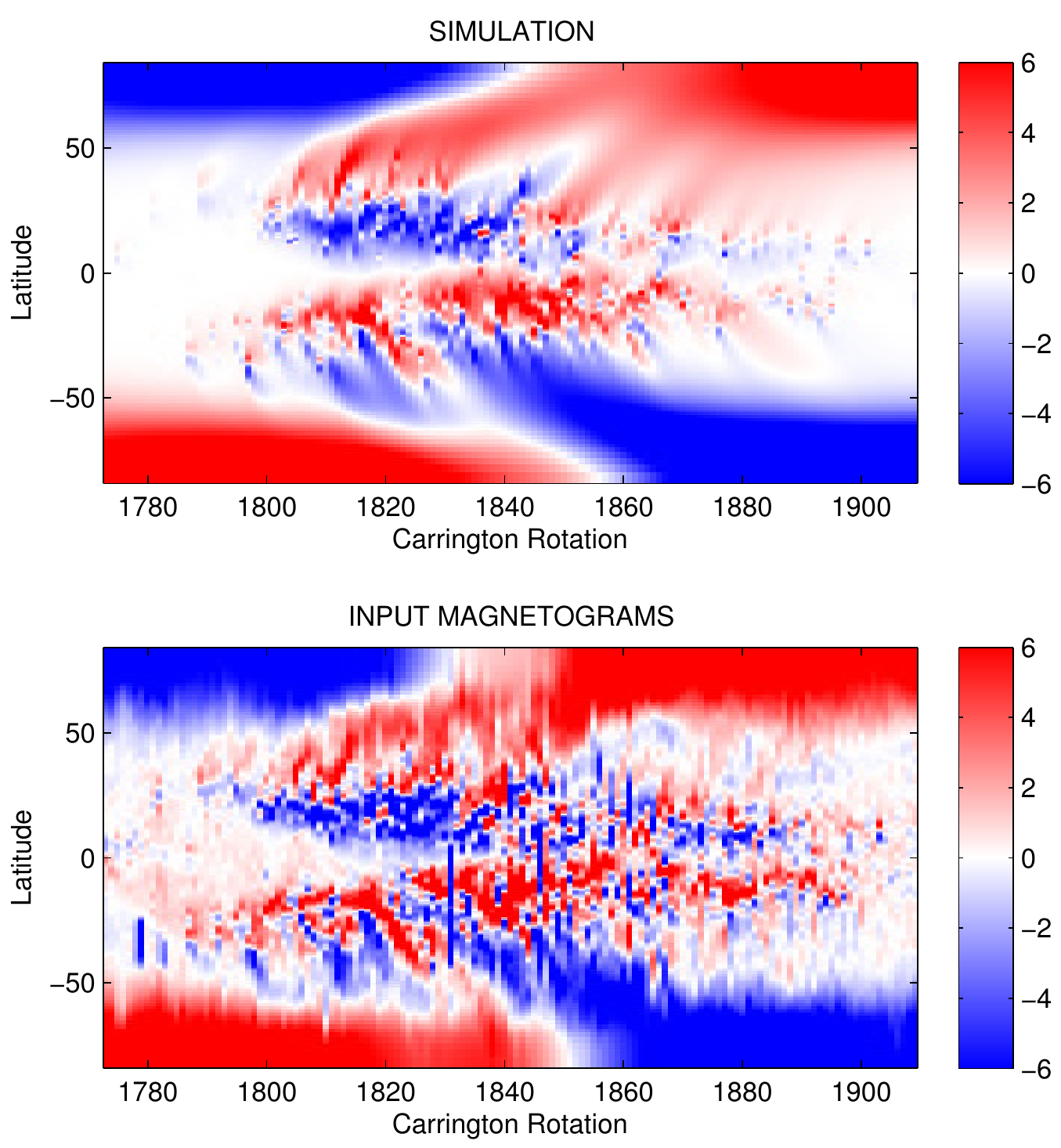}}
  \caption{Top: Butterfly diagram for the optimal parameter 4-set for the 2D model in Table \ref{table1big}(h). Bottom: Ground truth data for Cycle 22.}
  \label{2d4p22}
\end{figure}

Table \ref{table1big}(h) shows the optimization results for Cycle 22 (10\ts{th} March 1986--1\ts{st} June 1996). The fit is marginally worse than for Cycle 21, but optimal values for $\eta$ and $v_0$ remain within in plausible ranges. The optimal diffusion in this case increases to 506.2\,km$^2$\,s$^{-1}$, but is in better agreement with \citet{wangetal02b}. The optimal maximum velocity for Cycle 22 is even smaller than that of Cycle 21, further supporting the fact that a slower meridional flow results in a stronger polar field at cycle minimum, and explaining the high optimal maximum velocity for Cycle 23. \citet{vanballe98} performed SFT simulations for Cycle 22 with $\eta = 450$\,km$^2$\,s$^{-1}$ and $v_0 = 11$\,m\,s$^{-1}$ which produced polar field strength in agreement with observations. Again, these values are in accordance with ranges given in Table \ref{table1big}(h). 
\\\\
The ground truth data is shown in the bottom panel of Fig. \ref{2d4p22} and the simulated butterfly diagram is in the top panel of Fig. \ref{2d4p22}. The model has recreated polarity reversal much more successfully here, with only a slight delay in the north. Towards the end of the cycle there is a large build-up of positive flux and some weak, but visible, poleward surges in the northern hemisphere that have appeared in the simulation but are not observed in the real butterfly diagram.

\subsection{Cycle 24 (so far...)}

\begin{figure}
  \resizebox{\hsize}{!}{\includegraphics{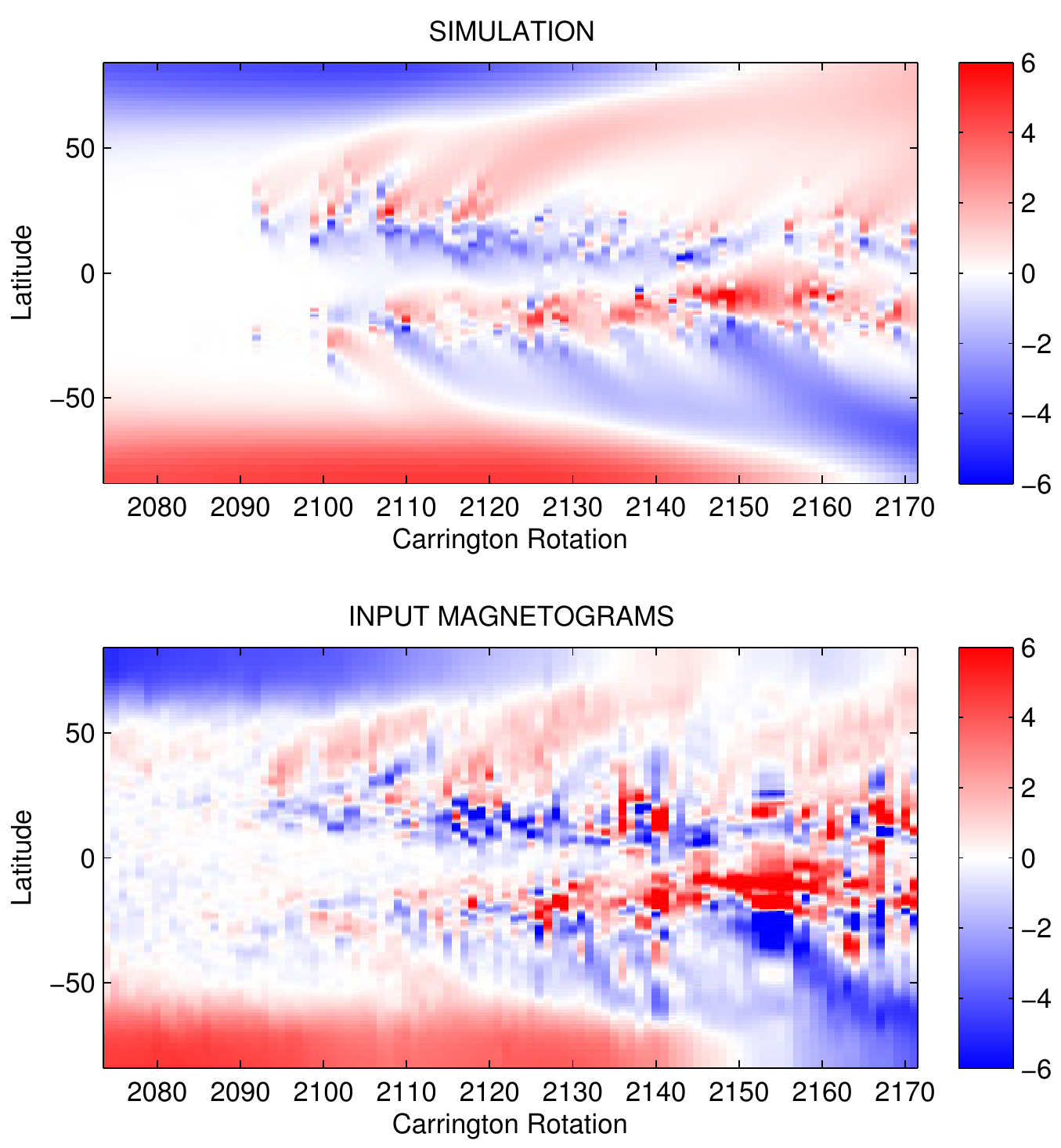}}
  \caption{Top: Butterfly diagram for the optimal parameter 4-set for the 2D model in Table \ref{table1big}(i). Bottom: Ground truth data for Cycle 24.}
  \label{2d4p24}
\end{figure}

Table \ref{table1big}(i) shows the results for the first half of Cycle 24 (3\ts{rd} August 2008--1\ts{st} Jan 2016). We obtain a much higher value of $\chi^{-2}$ for Cycle 24 compared to previous cycles, but we suspect that this is due to the relative ease of modelling only half a cycle as opposed to modelling long-term effects. The diffusivity $\eta = 454.6$\,km$^2$\,s$^{-1}$ is within viable ranges found in literature, though the maximum velocity is close to the lower prescribed bound. The initial polar field $B_0 = 4.2$\,G is lower than in previous cycles as the model needs to replicate the weak polar field at the Cycle 23/24 minimum. Acceptable ranges of parameters are generally broad, but performing the optimization on the full cycle in the next few years should tighten the upper and lower bounds. Indeed, when a similar optimization process is performed on half of Cycle 23, the acceptable ranges are found to be wider, though the shorter time period has a negligible effect on the specific optimal values.
\\\\
The interpolated Kitt Peak data is shown in the bottom panel of Fig. \ref{2d4p24} with the corresponding simulated butterfly diagram in the top panel of Fig. \ref{2d4p24}. Although a large portion of the cycle is yet to take place, there are still some notable features, such as the prominent leading-polarity region between 2100\,CR and 2110\,CR in the northern hemisphere. This region was the primary subject of \citet{yeates15}. Polar field reversal is slightly late in the simulated butterfly diagram; performing an optimization once the full cycle has completed might remedy this, though a region of negative polarity in the northern hemisphere at 2160\,CR may not correctly be reproduced, unless the data is corrected.\\

\noindent Including exponential decay in the model for Cycles 21, 22 and 24 produces optimal values of $\tau = 10.2 \in \left[3.1,32.0\right]$\,yr, $\tau = 7.6 \in \left[3.1,32.0\right]$\,yr and $\tau = 15.1 \in \left[2.5,32.0\right]$\,yr respectively. These are in better agreement with \citet{schrijver02} and \citet{lemerle}, indicating that the low optimal value for $\tau$ may only be necessary in Cycle 23 in order to successfully reconstruct the unusually weak polar fields at Cycle 23/24 minimum.

\section{Conclusions} \label{conclusions}

The aim of this paper was to use a genetic algorithm to find optimal parameters to be used in surface flux transport simulations, subsequently helping us understand the behaviour and interplay of the many physical processes on the Sun. We began by obtaining optimized parameter sets for a 1D SFT model for Cycle 23, both with and without a multiplicative tilt angle factor. From these simulations we obtained viable ranges for parameters. We found that these ranges and optimal solutions were in good agreement with results from previous studies and from observations. We also looked at the interaction of parameters, highlighting the positive correlations between the meridional velocity parameters $v_0$ and $p$, and exponential decay time $\tau$.
\\\\
We repeated the optimization process on a 2D assimilative model and found that optimum parameters were mostly within ranges of those from the 1D case, but distinct enough to suggest that the differences between models could be important. We also found an optimum value for the assimilation threshold, which was significantly greater than used previously by \citet{yeates15}. Qualitatively, the 2D model produced a more accurate butterfly diagram than the 1D model, particularly at the poles. We also included an exponential decay term in the 2D model which produced an optimal value of 4.5\,yr, which lies outside the acceptable range found in the 1D case and is in agreement with the values obtained by other authors. Including decay induced a decrease in the velocity parameters, but given that the acceptable range extended to the upper limits of exploration, its inclusion may not be necessary in the 2D model. There is the possibility that we did not model decay realistically, which could have led to a strong polar field. \bold{That the 2D model was able to give an acceptable match to the observed butterfly diagram and axial dipole moment without a decay term is evidence that it is superior to the 1D model, which was unable to do so with the corresponding optimal parameters. It suggests that the method of flux assimilation in the 2D model is superior to the insertion of idealized BMRs, as used both in the 1D model and in most other SFT models}.
\\\\
We were then able to compare the optimal meridional profiles from different regimes with observations made from feature tracking. The profiles from regimes (a), (c), and (d) were each almost completely within the range of observed flows, but the 1D optimal profile was faster than the average observed flow, while the 2D profile with decay included was too slow. The 2D profile without an extra decay term, however, best matched the average observed profile. Fixing the observed profile in both models resulted in varied success; the 2D model was able to accommodate the observations comfortably, whilst the 1D model saw a reduction in most parameters and a butterfly diagram containing an excess of flux in the transport regions.
\\\\
Finally, the optimization process was repeated using the 2D model for Cycles 21, 22, and 24, producing plausible results for Cycles 21 and 22; Cycle 24 may need more time to progress to capture the long-term effects of the cycle in the optimal parameters, particularly in narrowing some of the range of viable solutions, although an optimization run performed over the same period of time for Cycle 23 showed that the optimal parameters themselves are barely affected; it was just the ranges of acceptable values which widened due to fewer constraints. In order to predict the axial dipole moment at the Cycle 24/25 minimum and hence the amplitude and length of Cycle 25, randomly generated magnetic regions with properties based on empirical relations must be used to simulate the remainder of the cycle \citep[e.g.,][]{hathuptonpredict,cameronetal16}. Analysis of multiple cycles highlighted significant differences in meridional circulation speed, supporting the evidence for slower meridional flows during stronger cycles, and initial profile strength, supporting the proposed relationship between cycle strength and polar field strength at the preceding cycle minimum. Our multiple cycle analysis also highlighted cycle-dependence of the decay term $\tau$. At present, the best form and magnitude of such a decay term remain to be determined by the community. However, our results (and the others mentioned) do suggest that it can help to improve the match with observations, at least for Cycle 23. It is intriguing that it seems to be less important for the preceding cycles. This could either be because the decay is compensating for some other deficiency of the model that has changed in Cycle 23, such as the inability to reproduce the unusually weak polar field at the end of the cycle, or the radial diffusion of flux did really change from one cycle to the next, presumably due to some difference in the flows and magnetic field in the convection zone. This is an interesting subject for future study, but is beyond the scope of this paper where we consider only the surface. All optimization runs were performed with respect to a prescribed variance which was proportional to both latitude and $B_{obs}$, chosen to correspond to uncertainty in observational data. It should be noted that comparing fitness values is always with respect to the chosen error structure in this paper. For other studies of modelling Cycle 23 see, e.g., \citet{schrijver08,yeates10,yeates12,jiang13}.
\\\\
While the flexibility in the problem is beneficial in the respect that it allows more freedom, it can also have drawbacks. For example, the choice of fitness function is crucial to deciding which regime or parameter choice is `best' for each model, but depends entirely on what the user regards as important. \citet{lemerle} used a combination of $\chi ^2$ statistics which measured the differences between real and simulated time-latitude maps, axial dipoles and `transport regions' (latitudes $\pm 34$\textdegree to $\pm 54$\textdegree). These statistics were balanced equally in the final fitness function. Weighting could have been applied in favour of particular features, though it is not obvious how best to put this into practice. Alternatively, weighting could be applied to different sections of the map, i.e., active, transport and polar regions, to force the algorithm to return parameters which produce those specific regions more accurately. We chose a comparison between the real and simulated time-latitude maps, with an associated error structure, as we considered the general reproduction of the whole map to be foremost in importance.
\\\\
The adaptability of the 2D model provides a wide scope of possible future directions. One such direction is testing variability between different measuring instruments to ascertain whether inconsistent literature results could simply be due to the choice of observatory or satellite. This comes with the issue of either deciding on or computing an appropriate value for the assimilation threshold $BPAR$ for different datasets. Another future possibility that takes advantage of the model's assimilation technique is to optimize multiple cycles at the same time. We have shown that there exists variation in parameters between cycles, so a single optimal parameter set for more than one cycle would be unrealistic. An alternative method would be to treat each cycle separately, coupled only at each cycle minimum, where the final profile of the previous cycle becomes the initial profile of the next.
\\\\
Our methodology assumes a static meridional flow. The inclusion of a time-varying meridional flow in the optimization could significantly alter results, however parametrizing time-dependence without introducing too many parameters is not a trivial procedure. On the other hand, large-scale inflows towards active regions were first observed by \citet{gizon01}, and \citet{cameron12} proposed that these flows were at least partially responsible for variation of meridional flow over the solar cycle. Indeed, \citet{martinbelda} found that the inflows increased the effect of flux cancellation and also reduced the latitudinal separation of polarities, thereby decreasing the axial dipole moment contribution of a bipolar region. This process weakens the polar field in the same way that a time-dependent meridional flow can, and although we have not accounted for inflows in this study, it is an option under consideration for future work. An alternative method for reducing the polar field is using a flux-dependent diffusion parameter whereby the presence of a strong magnetic field quenches diffusion \citep[e.g.,][]{andres11}.
\\\\
In the near future we hope to use \texttt{PIKAIA} to optimize a kinematic 3D dynamo model \citep{kd3} using the results in this paper to constrain the surface evolution.

\begin{acknowledgements}

TW thanks Durham University Department of Mathematical Sciences for funding his PhD studentship. ARY thanks STFC for financial support. ARY and GJDP thank ISSI, Bern for supporting their collaboration as part of an International Team on global magnetic modelling. AMJ is very grateful to George Fisher and Stuart Bale for their support at the University of California, Berkeley, and Phil Scherrer for his support at Stanford University. This research is partly funded by the NASA Grand Challenge grant NNH13ZDA001N and NASA LWS grant NNX16AB77G. Data were acquired by SOLIS instruments operated by NISP/NSO/AURA/NSF. We are grateful to David Hathaway for also supplying data for this study and to Ian Vernon for his invaluable statistical advice\, \bold{as well as to the anonymous referee for useful suggestions}.

\end{acknowledgements}

\bibliographystyle{aa}
\bibliography{mybiblio}

\end{document}